\documentclass[prl,twocolumn,superscriptaddress,longbibliography]{revtex4-2}
\usepackage[colorlinks=true, citecolor=blue, urlcolor=blue]{hyperref}
\renewcommand{\section}[1]{{\par\it #1.---}\ignorespaces}
\usepackage{graphicx,amsmath,float}

\begin{document}
\title{Breakdown of Linearized Dispersion in Waveguide QED}
\author{H. Z. Shen}
\thanks{These authors contributed equally to this work.}
\affiliation{Center for Quantum Sciences and School of Physics, Northeast Normal University, Changchun 130024, China}
\author{Ze Wang}
\thanks{These authors contributed equally to this work.}
\affiliation{Center for Quantum Sciences and School of Physics, Northeast Normal University, Changchun 130024, China}
\author{X. X. Yi}
\email[Contact author: ]{yixx@nenu.edu.cn}
\affiliation{Center for Quantum Sciences and School of Physics, Northeast Normal University, Changchun 130024, China}
\author{Jun-Hong An}
\email[Contact author: ]{anjhong@lzu.edu.cn}
\affiliation{Key Laboratory of Quantum Theory and Applications of MoE, Lanzhou Center for Theoretical Physics, and Gansu Provincial Research Center for Basic Disciplines of Quantum Physics, Lanzhou University, Lanzhou 730000, China}
\date{\today}

\begin{abstract}
Exploring interactions between quantum emitters and guided photons, waveguide quantum electrodynamics (wQED) offers a versatile platform for quantum simulation and quantum interconnect. To simplify the quantized light-matter interactions, a linearized dispersion approximation (LDA) was widely used in wQED under the conditions of weak coupling and narrow photonic bandwidth. However, a growing number of strong-coupling wQED experiments have called the validity of LDA into question. Here, by investigating giant atoms coupled to a semi-infinite waveguide, we propose a principle to judge the validity of LDA. We find that the decoherence of the atoms exhibits rich dynamical behaviors determined by different numbers of bound states in and out of the continuum (BICs and BOCs) formed in the energy spectrum of the total atom-photon system. Based on this, we show that despite the LDA can partially preserve some of BICs, it completely erases BOCs. This causes the dominant role of LDA in the atomic dynamics to break down. Providing a physical insight into the validity of LDA, our results lay a firm foundation for exploring fundamental physics and its application frontier in strong-coupling wQED platforms.
\end{abstract}
\maketitle

\section{Introduction}
Waveguide quantum electrodynamics (wQED) provides a versatile platform to explore light-matter interactions and develop scalable quantum technologies \cite{Zheng2013,Awschalom2021,Tabares2023,Sheremet2023}. A key ingredient underlying its success is the linearized dispersion approximation (LDA) \cite{Shen2009,Shen2009a,Roy2017,Du2025}. With this clever trick, the complex photonic problems are converted into analytically tractable ones, which has enabled extensive advances in point- and giant-atom wQED systems \cite{Gustafsson2014,FornDiaz2017,Wen2019,Kannan2020,Almanakly2026,Yan2026}, as well as systems based on input-output descriptions \cite{Gardiner2004,Gough2009,Lalumiere2013,Caneva2015,Feng2021,Mirza2016a,Combes2017,Zhang2017,Kiilerich2019}. A broad range of quantum phenomena has been predicted and observed, including long-distance quantum-state transfer \cite{Stannigel2011,Vermersch2017,Diekmann2026a}, giant-atom interference effects \cite{FriskKockum2014,Carollo2020,Guo2020a,Wang2021,TerradasBrianso2022,Soro2022,Yin2025,Lue2026}, and retardation-induced quantum feedback \cite{Pichler2016,Calajo2019,Andersson2019,Roccati2024,BarahonaPascual2025}. The LDA has therefore played an important role in the development of wQED.

\begin{figure}[t]
\centering
\includegraphics[width=\columnwidth]{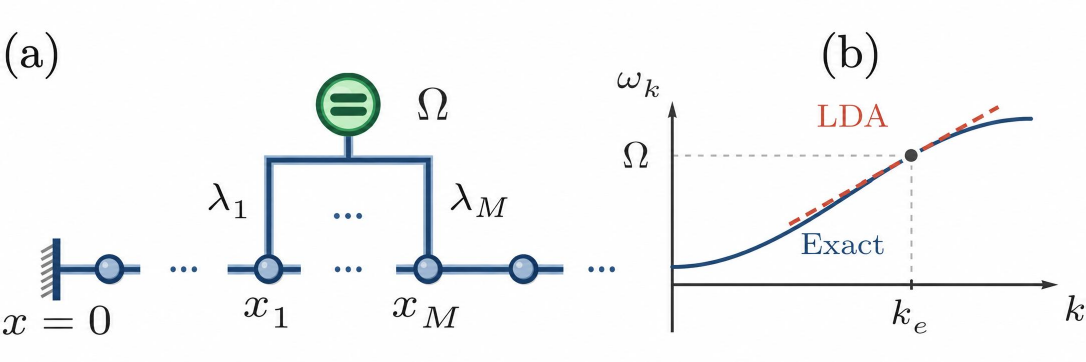}
\caption{(a) Schematic of a giant-atom waveguide QED setup. A two-level giant atom couples to a 1D semi-infinite waveguide at \(M\) points, where \(x_m\) and \(\lambda_m\) denote the position and coupling strength of the \(m\)th coupling point ($m=1,\cdots,M$), respectively. (b) Exact dispersion $\omega_k=\omega_0-2\lambda_0\cos k$ (blue solid curve) and its linearized version in Eq.~\eqref{Main:Eq8} (red dashed line).} \label{Fig1}
\end{figure}

The LDA expands the dispersion relation $\omega_{k}$ of electromagnetic field (EMF) in the waveguide to its first order around a resonant wave number $k_e$, i.e., $ \omega_{k} \approx \Omega + v (k - k_e) $ with $\Omega$ being the atomic transition frequency and $v$ being the group velocity \cite{Shen2009,Shen2009a,Roy2017,Tufarelli2013,Kockum2018,Guo2020}. Its validity is believed to be based on the assumption that only a narrow bandwidth in the vicinity of $\Omega$ has a significant contribution to the dynamics in the weak-coupling regime \cite{Shen2005,Walls2009,Koshino2012,Werra2013,Tufarelli2014,Schneider2016,Ramos2016,Fayard2021,Wang2024,Cilluffo2026}. The Markovian dynamics of a local quantum emitter coupled to a one-dimensional (1D) waveguide has even been proven to be exact under the LDA \cite{Shi2015}. The LDA has also been used to realize ``structured light-matter interactions'' by spatially designing the coupling sequence of giant atoms \cite{Wang2024}. However, it is reasonable to anticipate that this approximation would miss some essential physics, e.g., some transport features in the far-detuned regime \cite{Roy2011,Berndsen2023}. More critically, recent experiments in quantum engineering have accessed regimes of strong light-matter coupling, where band-edge effects significantly influence the dynamics. It goes well beyond the narrow-bandwidth assumption of LDA \cite{Liu2017,Mirhosseini2018,Sundaresan2019,Kim2021,Ferreira2021,Scigliuzzo2022,CastilloMoreno2025,Pakkiam2026}. This naturally challenges the LDA and raises a question of whether it remains valid in these regimes. A clear route toward a systematic understanding of when and why the LDA breaks down remains lacking.

In this Letter, we investigate the non-Markovian dynamics of giant atoms interacting with a semi-infinite waveguide and verify the breakdown of LDA. We establish a principle of ``energy-spectrum governing dynamics'' to judge its validity. Besides a continuous energy band, the energy spectrum of the atom-photon system supports bound states in and out of the continuum (BICs and BOCs), whose formation suppresses the atomic decoherence caused by the EMF. We show that, although the LDA preserves the BICs resonant to the atomic frequency, it completely removes BOCs and shifts the BICs off-resonant to the atomic frequency. The reason is that the LDA artificially transforms a finite-band photonic environment into a gapless one, thereby renormalizing the off-resonant BICs and leaving no space to support the BOCs. Therefore, whenever the BOC or the off-resonant BIC is present in the spectrum, the LDA fails to produce the actual dynamics. Identifying the origin of the breakdown of LDA to describe the diverse atomic dynamics, our results broaden the prospect of exploring the physics of wQED beyond the traditional narrow-bandwidth assumption.

\mbox{\textit{Energy-spectrum governed non-Markovian dynamics.-}}
--We consider a two-level giant atom coupled to a 1D semi-infinite waveguide with one end terminated by a perfect mirror at \(x=0\), see Fig.~\ref{Fig1}(a). The waveguide is modeled as a tight-binding coupled-resonator array. The momentum-space Hamiltonian is (\(\hbar \equiv 1\)) \cite{Tufarelli2013}
\begin{equation}
\hat H=\Omega\hat\sigma^\dagger\hat\sigma+\sum_k\omega_k\hat b_k^\dagger\hat b_k+\sum_{k}\sum_{m=1}^{M}(g_{km}\hat\sigma\hat b_k^\dagger+\mathrm{H.c.}),\label{Main:Eq1}
\end{equation}
where \(\hat\sigma\) is the transition operator from the excited state $|e\rangle$ to the ground state $|g\rangle$ of the atom with transition frequency \(\Omega\) and \(\hat b_k\) is the annihilation operator of the \(k\)th mode with frequency $\omega_k$ of the EMF in the waveguide. The coupling between the giant atom and the EMF mode of wave number \(k\) at the \(m\)th coupling point is \(g_{km}=\sqrt{2/(L+1)}\lambda_m\sin(kx_m)\), where \(L\) is the number of resonators and \(\lambda_m\) is the real-space coupling strength of the giant atom to the $m$th resonator site. The dispersion relation of the EMF \(\omega_k=\omega_0-2\lambda_0\cos k\) defines a finite photonic band with edges \(\omega_\pm=\omega_0\pm2\lambda_0\) \cite{SM}, where \(\omega_0\) denotes the frequency of resonators and \(\lambda_0\) is the photon hopping amplitude between adjacent resonators.

The Hamiltonian \(\hat H\) conserves the total excitation number \(\hat {\mathcal N}=\hat\sigma^\dagger\hat\sigma+\sum_k\hat b_k^\dagger\hat b_k\) due to \([\hat H,\hat {\mathcal N}]=0\). Therefore, under the initial condition \(|\Psi(0)\rangle = |e, \{0_k\}\rangle\), with \(|\{0_k\}\rangle\) denoting the vacuum state of the EMF in the waveguide, the evolved state of the total atom-waveguide system is expanded as \(|\Psi(t)\rangle=[c(t)\hat\sigma^\dagger+\sum_k d_k(t)\hat b_k^\dagger]|g, \{0_k\}\rangle\). Here, \(c(t)\) and \(d_k(t)\) are the excited-state probability amplitudes of the giant atom and the EMF. From the Schr\"odinger equation, we obtain 
\begin{equation}
\dot{c}(t)+i\Omega c(t)+\int_0^t d\tau G(t-\tau) c(\tau)=0,\label{Main:Eq2}
\end{equation}
where \(G(t)=\int_{\omega_-}^{\omega_+}d\omega J(\omega)e^{-i\omega t}\) is the correlation function of the EMF, with the spectral density defined by \(J(\omega)=\sum_{kmm'}g_{km}^*g_{km'}\delta(\omega-\omega_k)\). In the continuum limit \(L\to\infty\), this spectral density
takes the explicit form \(J(\omega)=\frac{2\rho(\omega)}{\pi}\sum_{mm'}\lambda_m\lambda_{m'}\sin(k_\omega x_m)\sin(k_\omega x_{m'})\), where
\(k_\omega=\arccos[(\omega_0-\omega)/(2\lambda_0)]\) is the wave number at frequency \(\omega\) and \(\rho(\omega)=[4\lambda_0^2-(\omega-\omega_0)^2]^{-1/2}\) is the density of states. The Laplace transform \(\widetilde F(s)=\int_{0}^{\infty}dt\,e^{-st}F(t)\) converts Eq.~\eqref{Main:Eq2} into \(\widetilde{c}(s)=[s+i\Omega+\widetilde{G}(s)]^{-1}\), where \(\widetilde{G}(s)=\int_{\omega_-}^{\omega_+}d\omega\, J(\omega)/(s+i\omega)\) is the Laplace transform of \(G(t)\). The analytical solution of \(c(t)\) is obtained by making the inverse Laplace transform to \(\widetilde{c}(s)\), which requires finding the pole of \(\widetilde{c}(s)\) via \cite{SM} \begin{equation}
Y(E) \equiv \Omega + \int_{\omega_-}^{\omega_+}d\omega\, \frac{J(\omega)}{E-\omega} = E, \label{Main:Eq3}
\end{equation}
with $s=-iE$. It is remarkable to find that the solutions of Eq. \eqref{Main:Eq3} are just the eigenenergies of Eq.~\eqref{Main:Eq1}. 
To be specific, we expand the eigenstate of \(\hat H\) as \(|\Phi\rangle=(\alpha\hat\sigma^\dagger +\sum_k\beta_k\hat b_k^\dagger)|g, \{0_k\}\rangle\). Substituting it into \(\hat H|\Phi\rangle = E|\Phi\rangle\) results in Eq. \eqref{Main:Eq3}. Equation~\eqref{Main:Eq3} has three types of roots. First, it has an infinite number of roots in the regime \(\omega_- < E <\omega_+\) as \(Y(E)\) is divergent for the ill-defined integrand. These roots form a continuous energy band. Second, \(Y(E)\) is a monotonically decreasing function of \(E\) when \(E < \omega_-\) or \(E > \omega_+\). Thus, Eq.~\eqref{Main:Eq3} could support an isolated root in these regimes provided \(Y(\omega_-) < \omega_- \) and \(Y(\omega_+) > \omega_+ \), respectively. 
Since these isolated states fall into the bandgap regimes, we call their corresponding eigenstates BOCs and denote their eigenenergies \(E_j^{\mathrm{boc}}\). It is straightforward to derive that a lower-branch BOC with \(E_j^{\mathrm{boc}}<\omega_-\) is present when
\begin{align}
\Omega-\omega_-&<\sum_{m,m'}\mathcal J_{mm'},\label{Main:Eq4}
\end{align} and an upper-branch BOC with \(E_j^{\mathrm{boc}}>\omega_+\) is present when
\begin{align}
\omega_+-\Omega&<\sum_{m,m'}(-1)^{x_m+x_{m'}}\mathcal J_{mm'},\label{Main:Eq5}
\end{align}
where \(\mathcal{J}_{mm'} = \lambda_m\lambda_{m'}\min(x_m,x_{m'})/\lambda_0\). Third, a removable singularity \(E^{\mathrm{bic}}_{j}\) determined by \(J(E^{\mathrm{bic}}_{j})=0\) could exist for \(Y(E)\) in the regime \(\omega_- <E<\omega_+\), which makes Eq. \eqref{Main:Eq3} form the BIC with eigenenergy \(E^{\mathrm{bic}}_{j}\) \cite{Chen2016,RaaholtIngelsten2024}.
The conditions for forming the BIC with
\begin{align}
\hspace{-0.2cm}E_j^{\mathrm{bic}}=\Omega+\sum_{m,m'}\wp_{mm'}^{j}[\sin(k_j d^{-}_{mm'})-\sin(k_j d^{+}_{mm'})],\label{Main:Eq6}
\end{align}
read $\sum_{m=1}^{M}\lambda_m\sin(k_j x_m)=0$, where \(\wp_{mm'}^j=\lambda_m\lambda_{m'}/v_j\), \(d_{mm'}^\pm=|x_m\pm x_{m'}|\). The wave number \(k_j\) is \(k_\omega\) evaluated at \(\omega=E_j^{\mathrm{bic}}\), and \(v_j=|\partial_k\omega_k|_{k=k_j}=2\lambda_0\sin k_j\). The BIC can be further classified into resonant BIC (rBIC) with \(E^{\mathrm{bic}}_{j} = \Omega\) and off-resonant BIC (oBIC) with \(E^{\mathrm{bic}}_{j} \neq \Omega\). Substituting the poles \(s=-iE_j^{\mathrm{boc/bic}}\) determined by Eq.~\eqref{Main:Eq3} into the inverse Laplace transform, we find that the bound-state energies \(E_j^{\mathrm{boc}}\) and \(E_j^{\mathrm{bic}}\) contribute nontrivial residues, while the band energies in the continuum contribute a branch cut, which tends to zero in the long-time limit due to the out-of-phase interference \cite{Liu2016,SanchezBurillo2017,Ji2022}. Thus, the analytical form of \(c(t)\) in the long-time limit is evaluated as
\begin{equation}
c(\infty)=\sum_{\nu = \mathrm{boc},\mathrm{bic}}\sum_j \operatorname*{Res}[\widetilde{c}(-iE_j^{\nu})]e^{-iE_{j}^{\nu}t}, \label{Main:Eq7}
\end{equation}
where \(\operatorname*{Res}[\widetilde c(-iE_j^\nu)]=[\partial_s\widetilde c(s)^{-1}|_{s=-iE_j^\nu}]^{-1}\) denotes the residue. From Eq.~\eqref{Main:Eq7}, we see that the dynamics of the giant atom characterized by \(c(t)\) is essentially determined by the energy-spectrum features of the total system \cite{Yang2025,Wu2026}. If the bound state is absent, we have the complete decoherence \(c(\infty)=0\), while if the bound state with energy \(E_{j}^{\nu}\) is formed, then we have the dissipationless dynamics in Eq.~\eqref{Main:Eq7}. This proves the one-to-one correspondence between the rich non-Markovian dynamical behaviors of open system and the energy-spectrum features of the total system.
Thus, the spectrum determined dynamics is established as a fundamental principle to unravel the physical mechanisms underlying the breakdown of LDA.

\section{Breakdown of LDA}
The LDA replaces the original dispersion relation \(\omega_k\) by a first-order expansion around the wave number \(k_e=\arccos[(\omega_0-\Omega)/(2\lambda_0)]\) as \cite{Shen2009,Shen2009a,Roy2017}
\begin{equation}
\omega_k^{\mathrm{LDA}}=\Omega+v(k-k_e), \label{Main:Eq8}
\end{equation}
where \(v=|\partial_k\omega_k|_{k=k_e}\) is the group velocity, see Fig.~\ref{Fig1}(b). Because the photonic dispersion can be engineered to produce exotic quantum dynamics \cite{Zhou2008a,GonzalezTudela2017,GonzalezTudela2017a,Bello2019}, replacing it by its linearized version raises the question of which dynamical features survive. Substituting Eq.~\eqref{Main:Eq8} into the correlation function $G(t)$ in Eq.~\eqref{Main:Eq2}, we obtain the LDA dynamical delay-differential equation \cite{SM}
\begin{equation}
\begin{aligned}
\dot c_{\mathrm{LDA}}(t)=&-i\Omega c_{\mathrm{LDA}}(t)+\sum_{\alpha=\pm}\sum_{m,m'}\alpha\Gamma_{mm'}^{\alpha} \\
&\times c_{\mathrm{LDA}}(t-\tau_{mm'}^\alpha)\Theta(\!t-\tau_{mm'}^\alpha\!),\label{Main:Eq9}
\end{aligned}
\end{equation}
where $\tau_{mm'}^{\alpha}=d^{\alpha}_{mm'}/v$ are the propagation delays, $\Gamma_{mm'}^{\alpha} = [\lambda_{m} \lambda_{m'} e^{i(k_ev-\Omega)\tau_{mm'}^{\alpha}}]/v$ denote the effective decay rates, and \(\Theta(t)\) is the Heaviside step function.
With the exact wave-number-dependent group velocity $|\partial_k\omega_k|$ replaced by a constant $v$, the delay-differential equation~\eqref{Main:Eq9} under the LDA greatly simplifies the analysis of photon propagation. The index $\alpha$ in Eq.~\eqref{Main:Eq9} explicitly distinguishes different photon-feedback processes. The direct transfer of a photon from the $m$th coupling point to the $m'$th coupling point without being reflected by the mirror corresponds to $\alpha=-$, while $\alpha=+$ means that the photon takes a mirror-reflected path \cite{Tufarelli2014,Li2024}.
The eigenenergy equation analogous to Eq.~\eqref{Main:Eq3} under the LDA reads
\begin{equation}
Y_{\mathrm{LDA}}(E)\equiv\Omega+\int_{-\infty}^{\infty}d\omega\frac{J_{\mathrm{LDA}}(\omega)}{E-\omega}=E.\label{Main:Eq10}
\end{equation}
The corresponding spectral density is \(J_{\mathrm{LDA}}(\omega)=\frac{2\rho(\Omega)}{\pi}\sum_{mm'}\lambda_m\lambda_{m'}\sin(k_\omega^{\scriptscriptstyle\mathrm{LDA}} x_m)\sin(k_\omega^{\scriptscriptstyle\mathrm{LDA}}x_{m'})\), with \(k_\omega^{\scriptscriptstyle\mathrm{LDA}}=k_e+(\omega-\Omega)/v\) being the wave number at frequency \(\omega\). Likewise, the pole equation of \(\widetilde{c}_{\mathrm{LDA}}(s)\) reduces to the LDA eigenenergy equation in Eq.~\eqref{Main:Eq10} by replacing \(s=-iE\). However, the use of LDA dramatically changes the energy spectrum of the total system. Artificially extending the integration limits \(\omega_-\) and \(\omega_+\) in Eq.~\eqref{Main:Eq3} to \(-\infty\) and \(+\infty\), the LDA transforms the original continuum band of the EMF into the one with infinite bandwidth. Consequently, this unbounded continuum leaves no space to support the formation of BOC. Thus, it is naturally expected that the LDA completely erases the BOC. To further check whether the LDA could capture the BICs, we derive that the BIC with eigenenergies
\begin{align}
\bar{E}^\text{bic}_j=\Omega+\sum_{m,m'} \bar{\wp}_{mm'} [\sin(\bar{k}_j d^{-}_{mm'}\!)-\sin(\bar{k}_j d^{+}_{mm'}\!)] \label{Main:Eq11}
\end{align} is present when $\sum_{m=1}^{M}\lambda_m\sin(\bar{k}_j x_m)=0$, where \(\bar{\wp}_{mm'}=\lambda_m\lambda_{m'}/v\) and
\(\bar k_j=\left.k_\omega^{\scriptscriptstyle\mathrm{LDA}}\right|_{\omega=\bar E_j^{\mathrm{bic}}}\). The LDA BIC conditions have a formal similarity to the exact ones in Eq. \eqref{Main:Eq6}. For the rBIC with \(E_j^{\mathrm{bic}}=\bar E^{\mathrm{bic}}_j=\Omega\), we have \(k_j=\bar k_j=k_e\) and \(v_j=v\). The formation conditions for the rBIC and its contribution to the long-time dynamics are therefore identical in the exact and LDA models. By contrast, the formation conditions for the oBIC differ between the two models because \(k_j\neq\bar k_j\) and \(v_j\neq v\) in general when \(E_j^{\mathrm{bic}}\neq\Omega\) and \(\bar E_j^{\mathrm{bic}}\neq\Omega\). Using the residue theorem, the long-time solution of \(c_{\mathrm{LDA}}(t)\) is obtained as
\begin{equation}
c_{\mathrm{LDA}}(\infty)=\sum_j \operatorname*{Res}[\widetilde{c}_{\mathrm{LDA}}(-i\bar{E}_j^{\mathrm{bic}})]e^{-i\bar{E}_{j}^{\mathrm{bic}}t}.
\label{Main:Eq12}
\end{equation}
Although the LDA preserves the rBIC contribution, it completely removes BOC and changes the oBIC. Therefore, according to our conclusion that the energy spectrum determines the dynamics, the LDA would break down as long as the BOC or oBIC is present in the exact energy spectrum.

\begin{figure}[t]
\centering
\includegraphics[width=\columnwidth]{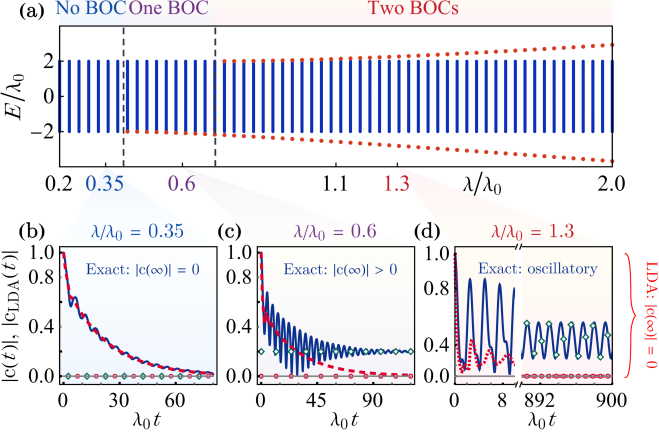}
\caption{BOC induced breakdown of LDA. (a) Exact energy spectrum of the total system via solving Eq.~\eqref{Main:Eq3}. The gray dashed lines mark the analytical BOC thresholds. (b)-(d) Evolution of probability-amplitude magnitudes for exact \(|c(t)|\) (blue solid curves) and LDA \(|c_{\mathrm{LDA}}(t)|\) (red dashed curves) for different values of $\lambda/\lambda_0$ in the presence of zero, one, and two BOCs. 
Long-time behaviors of \(|c(t)|\) (green diamonds) and \(|c_{\mathrm{LDA}}(t)|\) (red circles) are also shown. 
The parameters are \(x_1=1\), \(x_2=3\), \(\Omega=-\lambda_0\), \(\lambda\equiv\lambda_1=\lambda_2\), and \(\omega_0=0\).} \label{Main:Fig2}
\end{figure}

\section{Numerical results}
We choose two coupling points \(M=2\) and equal coupling strengths \(\lambda\equiv\lambda_1=\lambda_2\) as an example to illustrate our principle. First, we focus on a parameter regime with no BICs in both exact and linearized models. The exact energy spectrum of the total system in Fig.~\ref{Main:Fig2}(a) shows that the exact BOC branches marked by orange dots split the energy spectrum into three regions, without BOCs when \(\lambda\! <\! 0.44\lambda_0\), with one BOC when \(0.44\lambda_0\! <\! \lambda\! <\! 0.78\lambda_0\), and two BOCs when \(\lambda\! >\! 0.78\lambda_0\). By numerically solving Eqs.~\eqref{Main:Eq2} and \eqref{Main:Eq9}, we obtain the non-Markovian dynamics of the giant atom in the exact and LDA models for different values of \(\lambda\). In the absence of BOC at \(\lambda=0.35\lambda_0\), we observe \(\lvert c(t)\rvert\) decays to zero [see Fig.~\ref{Main:Fig2}(b)]. With one BOC formed when \(\lambda=0.6\lambda_0\), \(\lvert c(t)\rvert\) approaches a finite value [see Fig.~\ref{Main:Fig2}(c)]. With two BOCs formed when \(\lambda=1.3\lambda_0\), \(\lvert c(t)\rvert\) exhibits a lossless periodic oscillation [see Fig.~\ref{Main:Fig2}(d)]. By contrast, we find that \(\lvert c_{\mathrm{LDA}}(t)\rvert\) decays to zero irrespective of the value of \(\lambda\), which is the direct consequence of removal of BOCs under the LDA. The long-time behaviors of \(\lvert c(t)\rvert\) and \(\lvert c_{\mathrm{LDA}}(t)\rvert\) show good agreement with the analytical predictions of Eqs.~\eqref{Main:Eq7} and \eqref{Main:Eq12}. This confirms the deterministic role of the bound states in the dynamics of the giant atom.

\begin{figure}[t]
\centering
\includegraphics[width=\columnwidth]{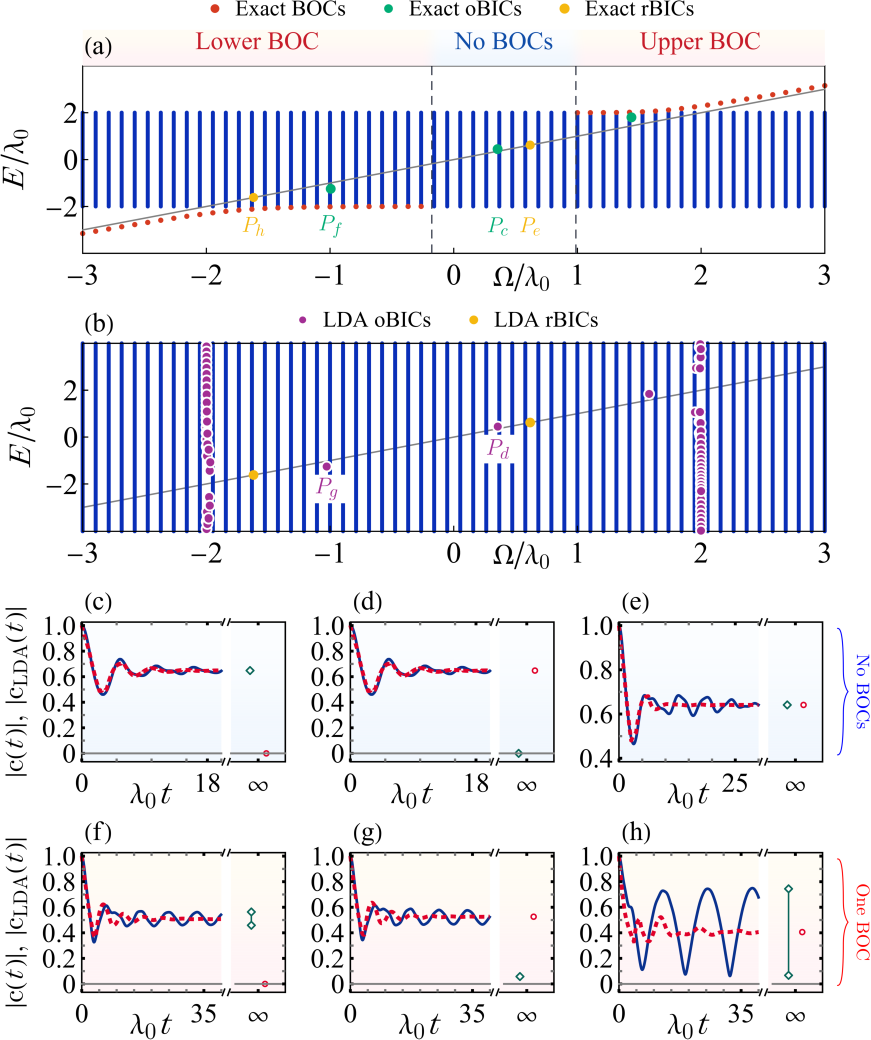} 
\caption{BIC induced breakdown of LDA. (a) Exact energy spectrum of the total system. The gray dashed lines mark the analytical BOC thresholds $\Omega = -0.18\lambda_0$ and $0.99\lambda_0$. (b) LDA energy spectrum of the total system. The gray solid line in (a) and (b) denotes $E=\Omega$. (c)-(h) Evolution of \(|c(t)|\) (blue solid curves) and LDA \(|c_{\mathrm{LDA}}(t)|\) (red dashed curves) at the marked points $P_c$-$P_h$. 
Long-time behaviors of \(\lvert c(t)\rvert\) and \(\lvert c_{\mathrm{LDA}}(t)\rvert\) are shown by green diamonds and red circles, respectively, with markers and stems indicating the long-time values and oscillation ranges.
Other parameters are \(x_1=1\), \(x_2=6\), \(\lambda_1=\lambda_2=0.45\lambda_0\), and \(\omega_0=0\).} \label{Main:Fig3}
\end{figure}

To reveal how the LDA reshapes BICs, we next compare the exact and LDA energy spectra in Fig.~\ref{Main:Fig3}. The exact energy spectrum in Fig.~\ref{Main:Fig3}(a) supports two branches of BOCs and a finite number of BICs, while the LDA energy spectrum in Fig.~\ref{Main:Fig3}(b) supports no BOCs and reveals complicated BICs. First, the rBICs (\(P_e\) and \(P_h\)) locate at the same positions in Figs.~\ref{Main:Fig3}(a) and \ref{Main:Fig3}(b), which verifies the identical rBIC condition under both the exact and LDA models. Second, two LDA oBICs \(P_d\) and \(P_g\) lie close to their exact counterparts \(P_c\) and \(P_f\), manifesting the reconstruction of oBICs by the LDA. Interestingly, an infinite sequence of oBICs due to the periodicity of Eq. \eqref{Main:Eq11} is uniquely present in the LDA spectrum, which have no counterparts in the exact spectrum and are artificially caused by the LDA. In the BOC-free regime, Figs.~\ref{Main:Fig3}(c)-\ref{Main:Fig3}(e) show the evolution of \(\lvert c(t)\rvert\) and \(\lvert c_{\mathrm{LDA}}(t)\rvert\) at different oBIC and rBIC conditions marked by \(P_c\), \(P_d\), and \(P_e\) in Fig.~\ref{Main:Fig3}(a) and \ref{Main:Fig3}(b). At the exact oBIC condition \(P_c\), \(\lvert c(t)\rvert\) tends to a nonzero value, while \(\lvert c_{\mathrm{LDA}}(t)\rvert\) decays to zero [Fig.~\ref{Main:Fig3}(c)]. The opposite behavior is found at the LDA oBIC condition \(P_d\) [Fig.~\ref{Main:Fig3}(d)]. 
In both cases of \(P_c\) and \(P_d\), \(\lvert c(t)\rvert\) and \(\lvert c_{\mathrm{LDA}}(t)\rvert\) are nearly indistinguishable at short times but qualitatively differ in the long-time limit. By contrast, at the rBIC condition \(P_e\), \(\lvert c(t)\rvert\) and \(\lvert c_{\mathrm{LDA}}(t)\rvert\) tend to the same long-time value [Fig.~\ref{Main:Fig3}(e)]. Figures~\ref{Main:Fig3}(f)-\ref{Main:Fig3}(h) show the results in the one-BOC regime at $P_f$, $P_g$, and $P_h$. We observe that although both of \(\lvert c(t)\rvert\) and \(\lvert c_{\mathrm{LDA}}(t)\rvert\) are governed by their respective energy-spectrum features, \(\lvert c_{\mathrm{LDA}}(t)\rvert\) cannot correctly describe the exact dynamics because it fails to capture the contribution of BOCs. 

\begin{figure}[t]
\centering
\includegraphics[width=\columnwidth]{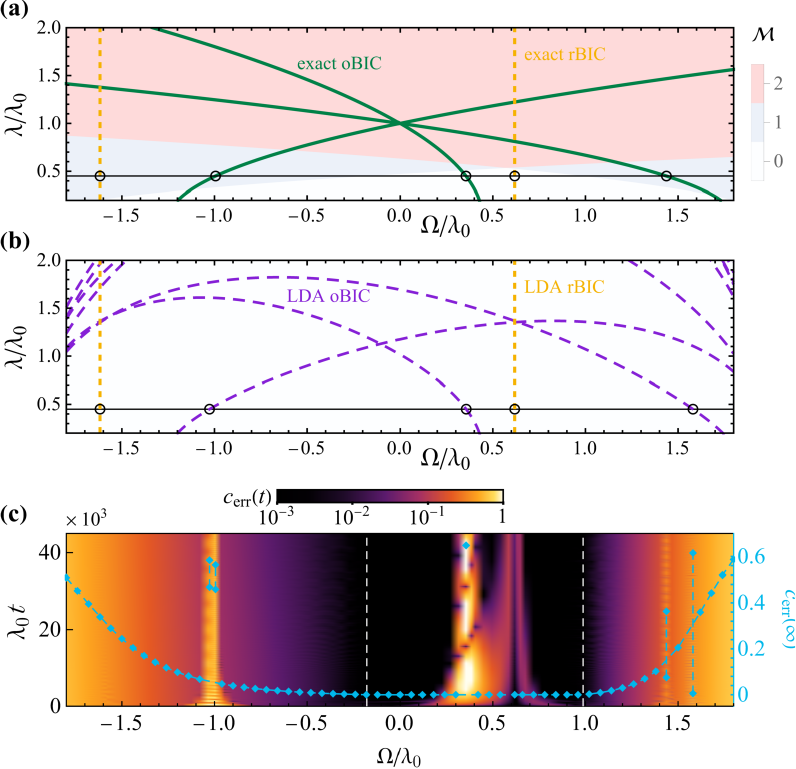}
\caption{(a) Bound-state phase diagram of the exact model. \(\mathcal M\) represents the number of BOCs. Green solid curves denote the exact oBIC branches, and yellow dashed lines mark the rBIC branches. (b) Bound-state phase diagram of the LDA model. Purple dashed curves denote the LDA oBIC branches and the rBIC branches are the same as those in (a). The black lines in (a) and (b) mark $\lambda/\lambda_0=0.45$, used in (c). The open circles indicate its intersections with the corresponding BIC branches. (c) Evolution of the probability-amplitude error \(c_{\mathrm{err}}(t)\equiv |c(t)-c_{\mathrm{LDA}}(t)|\) in different $\Omega$. Cyan diamonds and dashed stems indicate the long-time values and oscillation ranges of \(c_{\mathrm{err}}(\infty)\). White dashed vertical lines mark the exact BOC thresholds. Other parameters are the same as in Fig.~\ref{Main:Fig3}.} \label{Main:Fig4}
\end{figure}

To visualize the validity range of LDA, we plot in Figs.~\ref{Main:Fig4}(a) and \ref{Main:Fig4}(b) the bound-state phase diagrams of the exact and LDA models in the \((\Omega,\lambda)\) plane. The LDA remains valid in the white region excluding the exact and LDA oBIC branches in Fig.~\ref{Main:Fig4}(a). When the BOCs are present in the exact spectrum or the oBICs are present in either the exact or the LDA spectrum, the LDA fails to describe the dynamics of the giant atom. The dynamical consequences of the energy-spectrum differences between the exact and LDA models are quantified by the error \(c_{\mathrm{err}}(t)\equiv|c(t)-c_{\mathrm{LDA}}(t)|\). Figure~\ref{Main:Fig4}(c) shows the evolution of \(c_{\mathrm{err}}(t)\) for different \(\Omega\) when \(\lambda=0.45\lambda_0\). A clear difference occurs as long as the BOC is present in the exact energy spectrum or the oBIC is present in the exact or LDA spectrum. The long-time error agrees well with the analytical predictions of Eqs.~\eqref{Main:Eq7} and \eqref{Main:Eq12}, confirming that the validity of LDA is governed by the modification of energy spectrum.

\section{Discussion and conclusion}
The principle we propose to judge the validity of LDA is applicable to arbitrary multi-atom and multi-coupling-point configurations with the waveguide modeled by the tight-binding array of coupled resonators under both infinite and semi-infinite boundary conditions \cite{SM}. Long-range hopping and more general nonlinear dispersions could modify the bound-state conditions, while the dominant role of the energy spectrum in the dynamics is the same. Potential platforms for observing the predicted long-time signatures include superconducting circuits~\cite{Astafiev2010,Kannan2020}, coupled-cavity arrays~\cite{Notomi2008,Zhang2023a,Saxena2023}, and photonic-crystal waveguides~\cite{Arcari2014,Liu2017}.

In summary, we have established a general principle for judging the validity of LDA in wQED beyond the intuitive belief of a narrow-bandwidth assumption. We found that the breakdown of LDA in describing the non-Markovian dynamics of the giant atoms is essentially rooted in its modification to the energy spectrum of the total atom-EMF system. By analyzing the energy spectrum features, we revealed that although the LDA exactly captures the rBICs, it completely removes the BOCs and reconstructs the oBICs. Therefore, as long as either the BOC or the oBIC is present, it fails to describe the dynamics. Exhibiting rich giant-atom dynamics governed by the energy spectrum, our results shed light on the validity of LDA and expand the application frontier of strong-coupling wQED.

\section{Acknowledgments}
The work was supported by the National Natural Science Foundation of China (Grants No.~12575010, No.~92576202, and No.~12274064), the Science and Technology Development Plan Project of Jilin Province (Grant No.~20250102007JC), the Quantum Science and Technology-National Science and Technology Major Project (Grant No.~2023ZD0300904), and the Natural Science Foundation of Gansu Province (Grants No.~26RCKA011 and No.~25JRRA799).

\textit{Data Availability.---}
There are no publicly available research data or software supporting this manuscript. Requests for further information or data should be sent
to the authors.

\bibliography{SIGA}

\onecolumngrid
\begin{center}
    \large \textbf{End Matter}
\end{center}
\twocolumngrid
\setcounter{equation}{0}
\renewcommand{\theequation}{A\arabic{equation}}

\section{Multiple giant atoms}
We now generalize our study to $N$ two-level giant atoms coupling to a 1D semi-infinite waveguide with equal transition frequencies $\Omega_n=\Omega$ and $n=1,\ldots,N$. The Hamiltonian reads
\begin{equation}
\hat H_N=\Omega\sum_n\hat\sigma_n^\dagger\hat\sigma_n+\sum_k\omega_k\hat b_k^\dagger\hat b_k+\sum_{k,n}(g_{kn}\hat\sigma_n\hat b_k^\dagger+\mathrm{H.c.}), \label{EM:Eq1}
\end{equation}
where $\hat\sigma_n$ lowers the $n$th atom, and $g_{kn}=\sqrt{2/(L+1)}\sum_{m=1}^{M_n}\lambda_{nm}\sin(kx_{nm})$ is the coupling coefficient. Here, the \(n\)th giant atom couples to the waveguide at \(M_n\) positions \(x_{nm}\) with coupling strengths \(\lambda_{nm}\). The state of the total system at time \(t\) is expanded as
\(|\Psi_{N}(t)\rangle=[\sum_{n}c_{n}(t)\hat\sigma_{n}^\dagger+\sum_k d_k(t)\hat b_k^\dagger]|g^{\otimes N}, \{0_k\}\rangle\) 
under the initial condition \(c_1(0)=1\) and \(c_{n\neq 1}(0)=d_{k}(0)=0\). 
From the Schr\"odinger equation, we obtain the matrix-form dynamical equation of the giant atoms 
\begin{equation}
\dot{\boldsymbol c}(t)+i\Omega\boldsymbol c(t)+\int_0^t d\tau\,\boldsymbol G(t-\tau)\boldsymbol c(\tau)=\mathbf0,\label{EM:Eq2}
\end{equation}
where $\boldsymbol c(t)=[c_1 (t),\ldots,c_N (t)]^{\mathsf T}$ denotes the atomic probability-amplitude vector. The photonic correlation matrix is $\boldsymbol G(t)\equiv[G_{nn'}(t)]_{n,n'=1}^{N}=\int_{\omega_-}^{\omega_+}d\omega\,\boldsymbol J(\omega)e^{-i\omega t}$. For the spectral density matrix $\boldsymbol J(\omega)\equiv[J_{nn'}(\omega)]_{n,n'=1}^{N}$, we obtain $J_{nn'}(\omega)=\frac{2\rho(\omega)}{\pi}\sum_{m=1}^{M_n}\sum_{m'=1}^{M_{n'}}\lambda_{nm}\lambda_{n'm'}
\sin(k_\omega x_{nm})\sin(k_\omega x_{n'm'})$. Applying the Laplace transform to Eq.~\eqref{EM:Eq2} yields $\widetilde{\boldsymbol c}(s)=[s\mathbf I+i\Omega\mathbf I+\widetilde{\boldsymbol G}(s)]^{-1}\boldsymbol c(0)$, where $\mathbf I$ is the $N\times N$ identity matrix and $\widetilde{\boldsymbol G}(s)=\int_{\omega_-}^{\omega_+}d\omega\,\boldsymbol J(\omega)/(s+i\omega)$. The eigenstate of $\hat H_N$ in the single-excitation subspace can be written as \(|\Phi_N\rangle=(\sum_n\alpha_n\hat\sigma_n^\dagger+\sum_k\beta_k\hat b_k^\dagger)|g^{\otimes N},\{0_k\}\rangle\). Substituting this state into $\hat H_N|\Phi_N\rangle=E|\Phi_N\rangle$, we obtain the eigenenergy equation in matrix form
\begin{equation}
\boldsymbol Y(E)\boldsymbol\alpha\equiv[\Omega\mathbf I+\int_{\omega_-}^{\omega_+}d\omega\,\frac{\boldsymbol J(\omega)}{E-\omega}]\boldsymbol\alpha
=E\boldsymbol\alpha,\label{EM:Eq3}
\end{equation}
where $\boldsymbol\alpha=(\alpha_1,\ldots,\alpha_N)^{\mathsf T}$. By setting $s=-iE$, Eq.~\eqref{EM:Eq3} is identified as the pole equation of $\widetilde{\boldsymbol c}(s)$. The principle that the energy spectrum governs the dynamics thus extends to multiple giant atoms. Introducing
\(\boldsymbol\Delta(E)=\int_{\omega_-}^{\omega_+}d\omega\,\boldsymbol J(\omega)/(E-\omega)\), the lower- and upper-band BOC energies \(E_j^{\mathrm{boc}}<\omega_-\) and \(E_j^{\mathrm{boc}}>\omega_+\) are determined by \(\det[(E_j^{\mathrm{boc}}-\Omega)\mathbf I-\boldsymbol\Delta(E_j^{\mathrm{boc}})]=0\). Due to the monotonicity of the eigenvalue branches of \(E\mathbf I-\boldsymbol Y(E)\) outside the band, the corresponding existence conditions reduce to 
\begin{align}
\Delta_j^->\Omega-\omega_-~\text{and}~\Delta_j^+>\omega_+-\Omega,\label{EM:Eq4}
\end{align}
where \(\Delta_j^-\) and \(\Delta_j^+\) denote the eigenvalues of \(\boldsymbol\Delta^-\equiv-\lim_{E\to\omega_-^-}\boldsymbol\Delta(E)\) and \(\boldsymbol\Delta^+\equiv\lim_{E\to\omega_+^+}\boldsymbol\Delta(E)\) \cite{SM}. Counting the eigenvalues satisfying Eqs. \eqref{EM:Eq4} gives the numbers of BOCs below and above the band, respectively. Thus, $N$ giant atoms can support up to $N$ BOCs on each side of the band, compared with at most one for single giant atom. At the band edges, the matrix elements are calculated as \(\Delta_{nn'}^-=\lambda_0^{-1}\sum_{m\,m'}\lambda_{nm}\lambda_{n'm'}\min(x_{nm},x_{n'm'})\) and \(\Delta_{nn'}^+=\lambda_0^{-1}\sum_{m\,m'}(-1)^{x_{nm}+x_{n'm'}}\lambda_{nm}\lambda_{n'm'}\min(x_{nm},x_{n'm'})\). Inside the band, the collective BIC with energy \(E_j^{\mathrm{bic}}\) and atomic probability-amplitude vector \(\boldsymbol\alpha_j\) must satisfy 
\begin{align}
\boldsymbol J(E_j^{\mathrm{bic}})\boldsymbol\alpha_j&=\mathbf0,\label{EM:Eq5}\\
[(E_j^{\mathrm{bic}}-\Omega)\mathbf I-\boldsymbol\Delta(E_j^{\mathrm{bic}})]\boldsymbol\alpha_j&=\mathbf0.\label{EM:Eq6}
\end{align}
Compared with the single-giant-atom case, Eq.~\eqref{EM:Eq5} also allows destructive interference between radiation emitted by different atoms. At \(E=E_j^{\mathrm{bic}}\), evaluating \(\boldsymbol\Delta(E)\) gives \(\Delta_{nn'}(E_j^{\mathrm{bic}})=v_j^{-1}\sum_{m\,m'}\lambda_{nm}\lambda_{n'm'}[\sin(k_jd^-_{nm\,n'm'})-\sin(k_jd^+_{nm\,n'm'})]\), where \(d^\pm_{nm\,n'm'}=|x_{nm}\pm x_{n'm'}|\).

Due to the out-of-phase decoherence in the long-time limit, the inverse Laplace transform of Eq.~\eqref{EM:Eq2} gives 
\begin{equation}
\boldsymbol c(\infty)=\sum_{\nu=\mathrm{boc,bic}}\sum_j\boldsymbol R_j^\nu\boldsymbol c(0)e^{-iE_j^\nu t},\label{EM:Eq7}
\end{equation}
with $\boldsymbol R_j^\nu\equiv\operatorname*{Res}_{s=-iE_j^\nu}[s\mathbf I+i\Omega\mathbf I+\widetilde{\boldsymbol G}(s)]^{-1}$ being the residue matrix contributed by the bound-state energy $E_j^\nu$. It can be evaluated as $\boldsymbol R_j^\nu =\boldsymbol\alpha_j^\nu(\boldsymbol\alpha_j^\nu)^\dagger/\{(\boldsymbol\alpha_j^\nu)^\dagger[\mathbf I-\left.\partial_E\boldsymbol\Delta(E)\right|_{E=E_j^\nu}]\boldsymbol\alpha_j^\nu\}$ \cite{SM}, where $\boldsymbol\alpha_j^\nu$ satisfies $[E_j^\nu\mathbf I-\boldsymbol Y(E_j^\nu)] \boldsymbol\alpha_j^\nu=\mathbf0$. For multiple giant atoms, each bound-state contribution in Eq.~\eqref{EM:Eq7} depends on the initial atomic state $\boldsymbol c(0)$ and the target atom $n$. The matrix element $R_{j\,nn'}^\nu$ determines the contribution to atom $n$ from an excitation initially in atom $n'$.

\begin{figure}[t]
\centering
\includegraphics[width=\columnwidth]{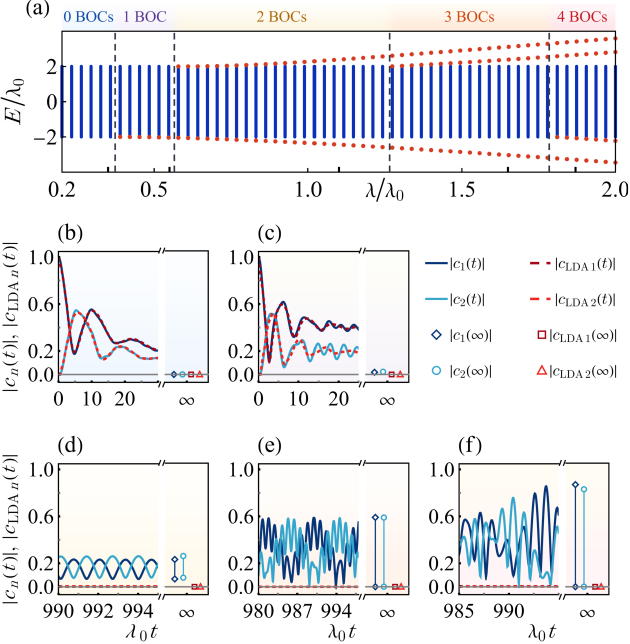}
\caption{(a) Exact energy spectrum. The dashed lines mark the analytical BOC thresholds. (b)-(f) \(|c_n(t)|\) and \(|c_{\mathrm{LDA}\,n}(t)|\) for \(\lambda/\lambda_0=0.35,0.55,1.10,1.65,1.87\), corresponding to \(0,1,2,3,4\) BOCs, respectively. Markers and stems indicate the long-time values and oscillation ranges. We use \(x_{11}=1\), \(x_{12}=4\), \(x_{21}=2\), \(x_{22}=5\), \(\lambda_{nm}\equiv\lambda\), \(\Omega=0.4\lambda_0\), and \(\omega_0=0\).}   \label{EM:Fig1}
\end{figure}

\section{Breakdown of LDA for multiple giant atoms}
Using the linearized dispersion in Eq.~\eqref{Main:Eq8}, the collective dynamics of the giant atoms under the LDA is
\begin{equation}
\begin{aligned}
&\dot c_{\mathrm{LDA}\,n}(t)=-i\Omega c_{\mathrm{LDA}\,n}(t)+\sum_{\alpha=\pm}\sum_{n',m,m'}\alpha\Gamma_{nm\,n'm'}^{\alpha}\\
&~~~~~~~~~~\times c_{\mathrm{LDA}\,n'}(t-\tau_{nm\,n'm'}^{\alpha})\Theta(t-\tau_{nm\,n'm'}^{\alpha}),
\end{aligned}\label{EM:Eq8}
\end{equation}
where \(\tau_{nm\,n'm'}^{\alpha}=d_{nm\,n'm'}^{\alpha}/v\) are the propagation delays and \(\Gamma_{nm\,n'm'}^{\alpha}=\lambda_{nm}\lambda_{n'm'}e^{i(k_ev-\Omega)\tau_{nm\,n'm'}^{\alpha}}/v\) denote the effective decay coefficients. 
The eigenenergy equation is
\begin{equation}
\boldsymbol Y_{\mathrm{LDA}}(E)\bar{\boldsymbol\alpha}
\equiv
[\Omega\mathbf I+\boldsymbol\Delta_{\mathrm{LDA}}(E)]
\bar{\boldsymbol\alpha}
=E\bar{\boldsymbol\alpha},
\label{EM:Eq9}
\end{equation}
where \(\bar{\boldsymbol\alpha}\) is the atomic probability-amplitude vector in the LDA eigenstate with energy \(E\) and the matrix \(\boldsymbol\Delta_{\mathrm{LDA}}(E)=\int_{-\infty}^{\infty}d\omega\,\boldsymbol J_{\mathrm{LDA}}(\omega)/(E-\omega)\). 
The LDA spectral density matrix $\boldsymbol J_{\mathrm{LDA}}(\omega)$ is obtained from $\boldsymbol J(\omega)$ by replacing $k_\omega$ with $k_\omega^{\mathrm{LDA}}$ and $\rho(\omega)$ with $\rho(\Omega)$. By setting $s=-iE$, Eq.~\eqref{EM:Eq9} is identical to the pole equation of $\widetilde{\boldsymbol c}_{\mathrm{LDA}}(s)$. The conditions for forming the \(j\)th LDA BIC with eigenenergies $\bar E_j^{\mathrm{bic}}$ are
\begin{align}
\boldsymbol J_{\mathrm{LDA}}(\bar E_j^{\mathrm{bic}})\bar{\boldsymbol\alpha}_j&=\mathbf0,\label{EM:Eq10}\\
\left[(\bar E_j^{\mathrm{bic}}-\Omega)\mathbf I-\boldsymbol\Delta_{\mathrm{LDA}}(\bar E_j^{\mathrm{bic}})\right]\bar{\boldsymbol\alpha}_j&=\mathbf0,\label{EM:Eq11}
\end{align}
where \(\bar{\boldsymbol\alpha}_j\) is the probability-amplitude vector of the LDA BIC and \(\Delta_{\mathrm{LDA}\,nn'}(\bar E_j^{\mathrm{bic}})=v^{-1}\sum_{m\,m'}\lambda_{nm}\lambda_{n'm'}[\sin(\bar k_jd^-_{nm\,n'm'})-\sin(\bar k_jd^+_{nm\,n'm'})]\). As in the single-atom case, the exact and LDA conditions are identical for collective rBICs, but generally differ for collective oBICs. Using the residue theorem, the collective LDA dynamics in the long-time limit is
\begin{equation}
\boldsymbol c_{\mathrm{LDA}}(\infty)=\sum_j\bar{\boldsymbol R}_j\boldsymbol c(0)e^{-i\bar E_j^{\mathrm{bic}}t},\label{EM:Eq12}
\end{equation}
where $\bar{\boldsymbol R}_j\equiv\operatorname*{Res}_{s=-i\bar E_j^{\mathrm{bic}}}[s\mathbf I+i\Omega\mathbf I+\widetilde{\boldsymbol G}_{\mathrm{LDA}}(s)]^{-1}$ is the residue matrix contributed by the LDA BIC energy $\bar E_j^{\mathrm{bic}}$.

\begin{figure}[b]
\centering
\includegraphics[width=\columnwidth]{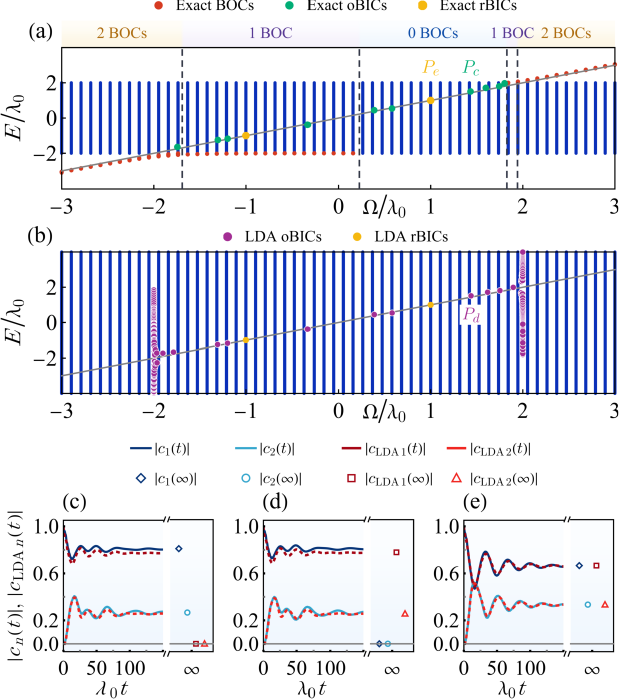}
\caption{(a) Exact and (b) LDA energy spectra. The dashed lines in (a) mark the analytical BOC thresholds. The gray solid line in (a) and (b) denotes $E=\Omega$, along with rBICs occur. (c)-(e) Evolution of \(|c_n(t)|\) (blue solid curves) and \(|c_{\mathrm{LDA}\,n}(t)|\) (red dashed curves) at the marked points $P_c$-$P_e$ in (a) and (b). We use \(x_{11}=3\), \(x_{12}=4\), \(x_{21}=7\), \(x_{22}=10\), \(\lambda_{nm}\equiv\lambda=0.24 \lambda_0\), and \(\omega_0=0\).}   \label{EM:Fig2}
\end{figure}

Figure~\ref{EM:Fig1} illustrates the collective BOC induced breakdown of the LDA. Figure~\ref{EM:Fig1}(a) shows the regions with zero to four exact BOCs. Figures~\ref{EM:Fig1}(b)-\ref{EM:Fig1}(f) show the corresponding dynamics at five coupling strengths. The exact \(|c_n(t)|\) decay to zero without BOCs, approach finite values with one BOC, and exhibit lossless oscillations with two to four BOCs. By contrast, the LDA \(|c_{\mathrm{LDA}\,n}(t)|\) decay to zero at all five couplings, verifying the removal of BOCs. Figure~\ref{EM:Fig2} illustrates the same spectral and dynamical distinction for two giant atoms. In the BOC-free regime, the exact and LDA dynamics differ in the long-time limit at the oBICs \(P_c\) and \(P_d\), but approach the same values at the rBIC \(P_e\). The results confirm that although the LDA successfully captures the rBICs, it fails to reproduce the oBICs and BOCs. Therefore, accompanying the formation of the oBICs and BOCs in the exact energy spectrum, its dominant role in describing the non-Markovian dynamics of the giant atoms breaks down. This gives a clear criterion to judge the performance of LDA in wQED.  

\end{document}


\title{Supplementary Material for ``Breakdown of Linearized Dispersion in Waveguide QED''}
\author{H. Z. Shen}
\thanks{These authors contributed equally to this work.}
\affiliation{Center for Quantum Sciences and School of Physics, Northeast Normal University, Changchun 130024, China}
\author{Ze Wang}
\thanks{These authors contributed equally to this work.}
\affiliation{Center for Quantum Sciences and School of Physics, Northeast Normal University, Changchun 130024, China}
\author{X. X. Yi}
\email[Contact author: ]{yixx@nenu.edu.cn}
\affiliation{Center for Quantum Sciences and School of Physics, Northeast Normal University, Changchun 130024, China}
\author{Jun-Hong An}
\email[Contact author: ]{anjhong@lzu.edu.cn}
\affiliation{Key Laboratory of Quantum Theory and Applications of MoE, Lanzhou Center for Theoretical Physics, and Gansu Provincial Research Center for Basic Disciplines of Quantum Physics, Lanzhou University, Lanzhou 730000, China}
\date{\today}
\maketitle

\section{Energy-spectrum governed non-Markovian dynamics}
\subsection{Derivation of Eq.~(2)}
For a two-level giant atom coupled to a 1D semi-infinite waveguide, the total real-space Hamiltonian is
\begin{equation}
\hat H=\Omega\hat\sigma^\dagger\hat\sigma+\omega_0\sum_x\hat b_x^\dagger\hat b_x-\lambda_0\sum_x(\hat b_x^\dagger\hat b_{x+1}+\mathrm{H.c.})+\sum_m\lambda_m(\hat\sigma\hat b_{x_m}^\dagger+\mathrm{H.c.}), \label{eq:SM_real_space_Hamiltonian}
\end{equation}
where \(\hat b_x\) annihilates a photon in the \(x\)th resonator of the array and \(x_m\) denotes the position of the \(m\)th coupling point between the giant atom and the waveguide. Applying the Fourier transform \(\hat b_x=\sqrt{2/(L+1)}\sum_k\sin(kx)\hat b_k\)~\cite{Longhi2006,Longhi2007}, the momentum-space Hamiltonian reads
\begin{equation}
\hat H=\Omega\hat\sigma^\dagger\hat\sigma+\sum_k\omega_k\hat b_k^\dagger\hat b_k+\sum_{k,m}
(g_{km}\hat\sigma\hat b_k^\dagger+\mathrm{H.c.}),\label{eq:SM_momentum_Hamiltonian}
\end{equation}
with the dispersion relation \(\omega_k=\omega_0-2\lambda_0\cos k\) and coupling coefficient \(g_{km}=\sqrt{2/(L+1)}\,\lambda_m\sin(kx_m)\).
The time-evolved state of the total system is expanded as \(|\Psi(t)\rangle=[c(t)\hat\sigma^\dagger+\sum_k d_k(t)\hat b_k^\dagger]|g,\{0_k\}\rangle\) under the initial condition \(|\Psi(0)\rangle=|e,\{0_k\}\rangle\).
From the Schrödinger equation \(i\partial_t|\Psi(t)\rangle=\hat H|\Psi(t)\rangle\), we obtain the following equations for the atomic and photonic amplitudes
\begin{align}
i\dot c(t)&=\Omega c(t)+\sum_{k,m}g_{km}^{*}d_k(t),
\label{eq:SM_atomic_amplitude_equation}\\
i\dot d_k(t)&=\omega_k d_k(t)+\sum_m g_{km}c(t).
\label{eq:SM_photonic_amplitude_equation}
\end{align}
Solving Eq.~\eqref{eq:SM_photonic_amplitude_equation}, we get
\begin{equation}
d_k(t)=-i\sum_m g_{km}\int_0^t d\tau\,e^{-i\omega_k(t-\tau)}c(\tau).
\label{TempEq1}
\end{equation}
Plugging Eq.~\eqref{TempEq1} back into Eq.~\eqref{eq:SM_atomic_amplitude_equation}, we obtain Eq.~(2) in the main text
\begin{equation}
\dot c(t)+i\Omega c(t)+\int_0^t d\tau\,G(t-\tau)c(\tau)=0,
\label{eq:SM_exact_memory_equation}
\end{equation}
where \(G(t)=\sum_{k,m,m'}g_{km}^{*}g_{km'}e^{-i\omega_k t}=\int_{\omega_-}^{\omega_+}d\omega\,J(\omega)e^{-i\omega t}\) is the environmental correlation function of the EMF, and \(J(\omega)\) is the spectral density, which can be calculated in the continuum limit \(\sum_k\to\frac{L+1}{\pi}\int_0^\pi dk\) as
\begin{align}
J(\omega)&=\sum_{k,m,m'}g_{km}^{*}g_{km'}\delta(\omega-\omega_k)
\notag\\
&=\frac{2}{\pi}\sum_{m,m'}\lambda_m\lambda_{m'}\int_0^\pi dk\,\sin(kx_m)\sin(kx_{m'})\delta(\omega-\omega_k)
\notag\\
&=\frac{2\rho(\omega)}{\pi}\sum_{m,m'}\lambda_m\lambda_{m'}\sin(k_\omega x_m)\sin(k_\omega x_{m'}),\label{eq:SM_exact_spectral_density}
\end{align}
where we have used the identity
$\delta[f(x)]=\sum_i\delta(x-x_i)/|f'(x_i)|$,
with $f(x_i)=0$ and $f'(x_i)\neq0$.

\subsection{Derivation of Eq.~(3)}
The Laplace transform \(\widetilde{c}(s)=\int_0^\infty dt\,c(t)e^{-st}\) converts Eq.~\eqref{eq:SM_exact_memory_equation} into \(\widetilde{c}(s)=[s+i\Omega+\widetilde{G}(s)]^{-1}\), where \(\widetilde{G}(s)=\int_{\omega_-}^{\omega_+}d\omega\,J(\omega)/(s+i\omega)\) is the Laplace transform of \(G(t)\). The analytical solution of \(c(t)\) is obtained by taking the inverse Laplace transform \(c(t)=(2\pi i)^{-1} \int_{\sigma-i\infty}^{\sigma+i\infty}ds\,\widetilde c(s)e^{st}\), where \(\sigma\) is chosen to be larger than all the poles of \(\widetilde{c}(s)\) and the poles of \(\widetilde c(s)\) are determined by the pole equation
\begin{equation}
s+i\Omega+\int_{\omega_-}^{\omega_+}d\omega\,\frac{J(\omega)}{s+i\omega}=0.
\label{eq:SM_pole_equation}
\end{equation}
By setting \(s=-iE\), we prove that the solutions of Eq.~\eqref{eq:SM_pole_equation} are just the eigenenergies of Eq.~\eqref{eq:SM_momentum_Hamiltonian} in the single-excitation subspace. 
To be specific, we assume that the eigenstate of \(\hat H\) is \(|\Phi\rangle=[\alpha\hat\sigma^\dagger+\sum_k\beta_k\hat b_k^\dagger]|g,\{0_k\}\rangle\). Then, imposing the eigenequation \(\hat H|\Phi\rangle=E|\Phi\rangle\), we get
\begin{align}
(E-\Omega)\alpha
&=\sum_{k,m}g_{km}^{*}\beta_k,
\label{eq:SM_eigen_atomic_equation}\\
(E-\omega_k)\beta_k&=\sum_m g_{km}\alpha.
\label{eq:SM_eigen_photonic_equation}
\end{align}
Solving Eq.~\eqref{eq:SM_eigen_photonic_equation} for \(\beta_k\) gives \(\beta_k=\alpha\sum_m g_{km}/(E-\omega_k)\). Substituting the result for \(\beta_k\) into Eq.~\eqref{eq:SM_eigen_atomic_equation}, we readily obtain Eq.~(3) in the main text
\begin{equation}
Y(E)\equiv \Omega+\int_{\omega_-}^{\omega_+}d\omega\,\frac{J(\omega)}{E-\omega}
=E.\label{eq:pole_equation}
\end{equation}

\subsection{Bound-state conditions}
In the regime \(E<\omega_-\) or \(E>\omega_+\), the integral in Eq.~\eqref{eq:pole_equation} can be evaluated as
\begin{align}
\int_{\omega_-}^{\omega_+}d\omega\,\frac{J(\omega)}{E-\omega}&=\frac{2}{\pi}\sum_{m,m'=1}^{M}\lambda_m\lambda_{m'}\int_0^\pi dk\,\frac{\sin(kx_m)\sin(kx_{m'})}{E-\omega_0+2\lambda_0\cos k}\nonumber\\
&=\frac{1}{\pi}\sum_{m,m'=1}^{M}\lambda_m\lambda_{m'}\int_0^\pi dk\,\frac{\cos[k(x_m-x_{m'})]-\cos[k(x_m+x_{m'})]}{E-\omega_0+2\lambda_0\cos k}\nonumber\\
&=\frac{\operatorname{sgn}(E-\omega_0)}{\sqrt{(E-\omega_0)^2-4\lambda_0^2}}\sum_{m,m'=1}^{M}\lambda_m\lambda_{m'}\big[\zeta_E^{|x_m-x_{m'}|}-\zeta_E^{x_m+x_{m'}}\big],\label{eq:SM_edge_shift_k_integral}
\end{align}
where \(\zeta_E=[\omega_0-E+\operatorname{sgn}(E-\omega_0)\sqrt{(E-\omega_0)^2-4\lambda_0^2}]/(2\lambda_0)\). Taking the lower- and upper-edge limits of Eq.~\eqref{eq:SM_edge_shift_k_integral} gives
\begin{align}
\lim_{E\to\omega_-^-}\int_{\omega_-}^{\omega_+}d\omega\,\frac{J(\omega)}{E-\omega}&=-\frac{1}{\lambda_0}\sum_{m,m'=1}^{M}\lambda_m\lambda_{m'}\min(x_m,x_{m'}),\label{eq:SM_lower_edge_shift}\\[1mm]
\lim_{E\to\omega_+^+}\int_{\omega_-}^{\omega_+}d\omega\,\frac{J(\omega)}{E-\omega}&=\frac{1}{\lambda_0}\sum_{m,m'=1}^{M}(-1)^{x_m+x_{m'}}\lambda_m\lambda_{m'}\min(x_m,x_{m'}).
\label{eq:SM_upper_edge_shift}
\end{align}
According to Eqs.~\eqref{eq:pole_equation}, \eqref{eq:SM_lower_edge_shift}, and \eqref{eq:SM_upper_edge_shift}, the conditions for the occurrence of BOCs with eigenenergies \(E_j^{\mathrm{boc}}\) are
\begin{align}
\Omega-\omega_-&<\frac{1}{\lambda_0}\sum_{m,m'=1}^{M}\lambda_m\lambda_{m'}\min(x_m,x_{m'}),\qquad E_j^{\mathrm{boc}}<\omega_-,\label{eq:SM_lower_BOC_condition}\\[1mm]
\omega_+-\Omega&<\frac{1}{\lambda_0}\sum_{m,m'=1}^{M}(-1)^{x_m+x_{m'}}\lambda_m\lambda_{m'}\min(x_m,x_{m'}),\qquad E_j^{\mathrm{boc}}>\omega_+.\label{eq:SM_upper_BOC_condition}
\end{align}
In the regime \(\omega_-<E<\omega_+\), the integral in Eq.~\eqref{eq:pole_equation} is evaluated as
\begin{align}
\mathcal P\int_{\omega_-}^{\omega_+}d\omega\,\frac{J(\omega)}{E-\omega}&=\frac{2}{\pi}\sum_{m,m'=1}^{M}\lambda_m\lambda_{m'}\mathcal P\int_0^\pi dk\,\frac{\sin(kx_m)\sin(kx_{m'})}{E-\omega_0+2\lambda_0\cos k}\nonumber\\
&=\frac{1}{2\pi\lambda_0}\sum_{m,m'=1}^{M}\lambda_m\lambda_{m'}\mathcal P\int_0^\pi dk\,\frac{\cos(kd_{mm'}^-)-\cos(kd_{mm'}^+)}{\cos k-\cos k_E}\nonumber\\
&=\frac{1}{v_E}\sum_{m,m'=1}^{M}\lambda_m\lambda_{m'}[\sin(k_Ed_{mm'}^-)-\sin(k_Ed_{mm'}^+)],\label{eq:SM_BIC_Lamb_shift}
\end{align}
where \(\mathcal P\) denotes the Cauchy principal value, and \(k_E=k_{\omega}|_{\omega=E}\), \(v_E=|\partial_k \omega_k|_{k=k_E}\). The BICs with energies \(\omega_-<E_j^{\mathrm{bic}}<\omega_+\) must satisfy \(J(E_j^{\mathrm{bic}})=0\) and \(Y(E_j^{\mathrm{bic}})=E_j^{\mathrm{bic}}\). According to Eqs.~\eqref{eq:SM_exact_spectral_density}, \eqref{eq:pole_equation}, and \eqref{eq:SM_BIC_Lamb_shift}, the BIC conditions are given by
\begin{gather}
\sum_{m=1}^{M}\lambda_m\sin(k_jx_m)=0,\label{eq:SM_BIC_dark_condition}\\[1mm]
E_j^{\mathrm{bic}}=\Omega+\frac{1}{v_j}\sum_{m,m'=1}^{M}\lambda_m\lambda_{m'}[\sin(k_jd_{mm'}^-)-\sin(k_jd_{mm'}^+)],\label{eq:SM_BIC_energy_condition}
\end{gather}
where \(k_j=k_E|_{E=E_j^{\mathrm{bic}}}\) and \(v_j=v_E|_{E=E_j^{\mathrm{bic}}}\). Under the condition of \(M=1\), Eqs.~\eqref{eq:SM_BIC_dark_condition} and \eqref{eq:SM_BIC_energy_condition} reduce to \(\sin(k_jx_1)=0\) and \(E_j^{\mathrm{bic}}=\Omega\), recovering the BIC condition for a single discrete state coupled to a semi-infinite lattice~\cite{Longhi2007}.

\begin{figure}[t]
\centering
\includegraphics[width=0.52\columnwidth]{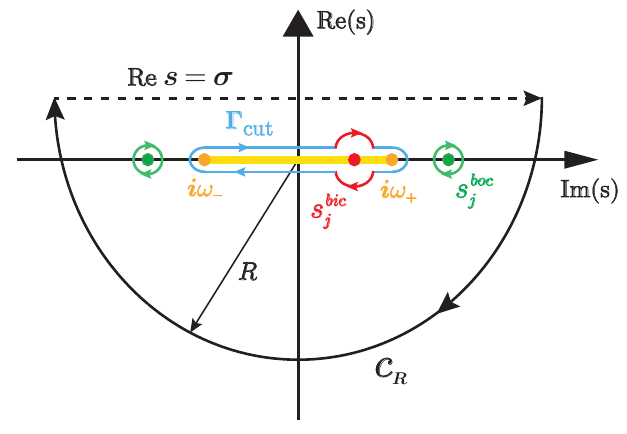}
\caption{Path of contour integration in the complex \(s\) plane for calculating the inverse Laplace transform of \(\widetilde c(s)\). Without loss of generality, we set \(\omega_0=0\). The dashed line denotes the Bromwich path at \(\operatorname{Re}s=\sigma\), and \(\mathcal C_R\) denotes the large closing arc. The blue path \(\Gamma_{\mathrm{cut}}\) denotes the branch cut. The green points \(s_j^{\mathrm{boc}}\) denote BOC poles on the imaginary axis external to the branch cut. The red points \(s_j^{\mathrm{bic}}\) denote the BIC poles on the imaginary axis internal to the branch cut.} \label{fig:SM_Bromwich_contour}
\end{figure}

\subsection{Derivation of Eq.~(7)}
Figure~\ref{fig:SM_Bromwich_contour} shows the path of the contour integration to evaluate the inverse Laplace transform of \(\widetilde c(s)\) and to obtain the analytical solution \(c(t)\) of Eq.~\eqref{eq:SM_exact_memory_equation}. According to the residue theorem, the contour deformation in Fig.~\ref{fig:SM_Bromwich_contour} gives
\begin{equation}
\lim_{R\to\infty}[\int_{\sigma-iR}^{\sigma+iR}+\int_{\mathcal C_R}]ds\,e^{st}\widetilde c(s)=2\pi i\sum_{\nu=\mathrm{boc,bic}}\sum_j\operatorname*{Res}_{s=-iE_j^\nu}[e^{st}\widetilde c(s)]+\int_{\Gamma_{\mathrm{cut}}}ds\,e^{st}\widetilde c(s).
\label{eq:SM_residue_contour}
\end{equation}
where \(\operatorname*{Res}_{s=-iE_j^\nu}[e^{st}\widetilde c(s)]\) denotes the residue contributed by the \(j\)th bound state of type \(\nu=\mathrm{boc,bic}\). From Jordan's lemma, the integral along the large arc \(\mathcal C_R\) vanishes as \(R\to\infty\). We then have
\begin{equation}
c(t)=\sum_{\nu=\mathrm{boc,bic}}\sum_j\operatorname*{Res}_{s=-iE_j^\nu}[e^{st}\widetilde c(s)]+\frac{1}{2\pi i}\int_{\Gamma_{\mathrm{cut}}}ds\,e^{st}\widetilde c(s).\label{eq:SM_pole_cut_decomposition}
\end{equation}
The integral along the branch cut in Eq.~\eqref{eq:SM_pole_cut_decomposition} tends to zero in the long-time limit due to the out-of-phase decoherence of the components oscillating at continuously different frequencies~\cite{SanchezBurillo2017}. In the long-time limit, we have
\begin{equation}
c(\infty)=\sum_{\nu=\mathrm{boc,bic}}\sum_j\operatorname*{Res}[\widetilde c(-iE_j^\nu)]e^{-iE_j^\nu t},
\label{eq:SM_long_time_bound_state_solution}
\end{equation}
where \(\operatorname*{Res}[\widetilde c(-iE_j^\nu)]=[\partial_s\widetilde c(s)^{-1}|_{s=-iE_j^\nu}]^{-1}\), which can be evaluated using L'H\^{o}pital's rule.

\section{Breakdown of LDA}
\subsection{Derivation of Eq.~(9)}
The LDA dynamics is obtained simply by replacing the exact dispersion \(\omega_k\) in the environmental correlation function \(G(t)\) entering Eq.~\eqref{eq:SM_exact_memory_equation} with its LDA version \(\omega_k^{\mathrm{LDA}}\). This turns \(J(\omega)\) into \(J_{\mathrm{LDA}}(\omega)\), which is evaluated as
\begin{align}
J_{\mathrm{LDA}}(\omega)&=\sum_{k,m,m'}g_{km}^{*}g_{km'}\delta(\omega-\omega_k^{\mathrm{LDA}})\notag\\[-1mm]&=\frac{2}{\pi}\sum_{m,m'}\lambda_m\lambda_{m'}\int_{-\infty}^{\infty}dq\,\sin[(k_e+q)x_m]\sin[(k_e+q)x_{m'}]\delta(\omega-\Omega-vq)\notag\\
&=\frac{2}{\pi v}\sum_{m,m'}\lambda_m\lambda_{m'}\sin(k_\omega^{\mathrm{LDA}}x_m)\sin(k_\omega^{\mathrm{LDA}}x_{m'}),\label{eq:SM_LDA_spectral_density}
\end{align}
where \(k_\omega^{\mathrm{LDA}}=k_e+(\omega-\Omega)/v\) and \(q=k-k_e\). The LDA environmental correlation function \(G_{\mathrm{LDA}}(t)\) is obtained from \(J_{\mathrm{LDA}}(\omega)\) as
\begin{align}
G_{\mathrm{LDA}}(t)&=\int_{-\infty}^{\infty}d\omega\,J_{\mathrm{LDA}}(\omega)e^{-i\omega t}\notag\\
&=\frac{1}{\pi v}\sum_{m,m'}\lambda_m\lambda_{m'}\int_{-\infty}^{\infty}d\omega\,e^{-i\omega t}[\cos(k_\omega^{\mathrm{LDA}}d_{mm'}^-)-\cos(k_\omega^{\mathrm{LDA}}d_{mm'}^+)]\notag\\
&=\sum_{m,m'}[\Gamma_{mm'}^-\delta(t-\tau_{mm'}^-)-\Gamma_{mm'}^+\delta(t-\tau_{mm'}^+)].\label{eq:SM_LDA_correlation}
\end{align} 
Plugging Eq.~\eqref{eq:SM_LDA_correlation} into Eq.~\eqref{eq:SM_exact_memory_equation}, we obtain the LDA dynamical delay-differential equation (9) in the main text
\begin{align}
\dot c_{\mathrm{LDA}}(t)=&-i\Omega c_{\mathrm{LDA}}(t)\notag\\
&-\sum_{m,m'}^{M}\Gamma_{mm'}^-\int_0^t dt'\,\delta(t-t'-\tau_{mm'}^-)c_{\mathrm{LDA}}(t')\notag\\
&+\sum_{m,m'}^{M}\Gamma_{mm'}^+\int_0^t dt'\,\delta(t-t'-\tau_{mm'}^+)c_{\mathrm{LDA}}(t')\notag\\
=&-i\Omega c_{\mathrm{LDA}}(t)\notag\\
&-\sum_{m,m'}^{M}\Gamma_{mm'}^-c_{\mathrm{LDA}}(t-\tau_{mm'}^-)\Theta(t-\tau_{mm'}^-)\notag\\
&+\sum_{m,m'}^{M}\Gamma_{mm'}^+c_{\mathrm{LDA}}(t-\tau_{mm'}^+)\Theta(t-\tau_{mm'}^+),\label{eq:SM_LDA_delay_equation}
\end{align}
where \(\Theta(t)\) is the Heaviside step function with \(\Theta(t)=1\) for \(t\geqslant0\) and \(\Theta(t)=0\) for \(t<0\).

\subsection{Derivation of Eq.~(10)}
Taking the Laplace transform of Eq.~\eqref{eq:SM_LDA_delay_equation} with \(c_{\mathrm{LDA}}(0)=1\), we obtain the pole equation of \(\widetilde c_{\mathrm{LDA}}(s)\)
\begin{equation}
s+i\Omega+\sum_{m,m'}\Gamma_{mm'}^-e^{-s\tau_{mm'}^-}-\sum_{m,m'}\Gamma_{mm'}^+e^{-s\tau_{mm'}^+}=0.\label{eq:SM_LDA_Laplace}
\end{equation}
Replacing \(s=-iE\) in Eq.~\eqref{eq:SM_LDA_Laplace} and imposing the LDA BIC conditions (Eqs.~\eqref{eq:SM_LDA_BIC_dark_condition} and \eqref{eq:SM_LDA_BIC_energy_condition}) to Eq.~\eqref{eq:SM_LDA_Laplace} gives
\begin{equation}
E-\Omega-\sum_{m,m'}\frac{\lambda_m\lambda_{m'}}{v}(\sin(k_E^{\mathrm{LDA}}d_{mm'}^-)-\sin(k_E^{\mathrm{LDA}}d_{mm'}^+)) = 0.\label{eq:SM_LDA_pole_equation}
\end{equation}
From the definition of the LDA in Eq.~(8) of the main text, the LDA energy transcendental function in Eq.~(10) is obtained by replacing \(J(\omega)\) with \(J_{\mathrm{LDA}}(\omega)\) and extending the integration limits to \(-\infty\) and \(+\infty\) in Eq.~\eqref{eq:pole_equation}
\begin{equation}
Y_{\mathrm{LDA}}(E) \equiv \Omega+\int_{-\infty}^{\infty}d\omega\,\frac{J_{\mathrm{LDA}}(\omega)}{E-\omega}=E.\label{eq:SM_LDA_energy_equation}
\end{equation}
The integral in Eq.~\eqref{eq:SM_LDA_energy_equation} can be analytically calculated as
\begin{align}
\mathcal P\int_{-\infty}^{\infty}d\omega\,\frac{J_{\mathrm{LDA}}(\omega)}{E-\omega}
&=\frac{2}{\pi v}\sum_{m,m'}\lambda_m\lambda_{m'}\mathcal P\int_{-\infty}^{\infty}dk\,\frac{\sin(kx_m)\sin(kx_{m'})}{k_E^{\mathrm{LDA}}-k}\nonumber\\
&=\frac{1}{\pi v}\sum_{m,m'}\lambda_m\lambda_{m'}\mathcal P\int_{-\infty}^{\infty}dk\,\frac{\cos(kd_{mm'}^-)-\cos(kd_{mm'}^+)}{k_E^{\mathrm{LDA}}-k}\nonumber\\
&=\sum_{m,m'}\frac{\lambda_m\lambda_{m'}}{v}[\sin(k_E^{\mathrm{LDA}}d_{mm'}^-)-\sin(k_E^{\mathrm{LDA}}d_{mm'}^+)],\label{eq:SM_LDA_Lamb_shift}
\end{align}
where \(k_E^{\mathrm{LDA}}=k_{\omega}^{\mathrm{LDA}}|_{\omega=E}\). 
It is easy to see that Eq.~\eqref{eq:SM_LDA_pole_equation} is just Eq.~\eqref{eq:SM_LDA_energy_equation}.

\subsection{LDA bound-state conditions} \label{sec:LDA_bound_states}
Since the LDA completely erases the BOC, we derive the LDA BIC conditions here.
Similarly to the derivation of the exact model BIC, the LDA BIC with energy \(\bar E_j^{\mathrm{bic}}\) also must satisfy \(J_{\mathrm{LDA}}(\bar E_j^{\mathrm{bic}})=0\) and \(Y_{\mathrm{LDA}}(\bar E_j^{\mathrm{bic}})=\bar E_j^{\mathrm{bic}}\). According to Eqs.~\eqref{eq:SM_LDA_spectral_density}, \eqref{eq:SM_LDA_energy_equation}, and \eqref{eq:SM_LDA_Lamb_shift}, the LDA BIC conditions are
\begin{gather}
\sum_{m=1}^{M}\lambda_m\sin(\bar k_j x_m)=0,\label{eq:SM_LDA_BIC_dark_condition}\\[1mm]\bar E_j^{\mathrm{bic}}=\Omega+\sum_{m,m'=1}^{M}\frac{\lambda_m\lambda_{m'}}{v}[\sin(\bar k_j d_{mm'}^-)-\sin(\bar k_j d_{mm'}^+)],\label{eq:SM_LDA_BIC_energy_condition}
\end{gather}
where \(\bar k_j=k_{\omega}^{\mathrm{LDA}}|_{\omega=\bar E_j^{\mathrm{bic}}}\). Note that the LDA BIC energy \(\bar E_j^{\mathrm{bic}}\) is not restricted in its range of values unlike the exact BIC.

\subsection{Derivation of Eq.~(12)}
Since $\widetilde c_{\mathrm{LDA}}(s)$ is meromorphic, no branch-cut contribution arises and only the LDA BIC contributions remain in the long-time limit
\begin{equation}
c_{\mathrm{LDA}}(\infty)=\sum_j\operatorname*{Res}[\widetilde c_{\mathrm{LDA}}(-i\bar E_j^{\mathrm{bic}})]e^{-i\bar E_j^{\mathrm{bic}}t},\label{eq:SM_LDA_long_time_solution}
\end{equation}
where \(\operatorname*{Res}[\widetilde c_{\mathrm{LDA}}(-i\bar E_j^{\mathrm{bic}})]=[\partial_s\widetilde c_{\mathrm{LDA}}(s)^{-1}|_{s=-i\bar E_j^{\mathrm{bic}}}]^{-1}\). Importantly, not only the formation condition for the rBIC is identical in the exact and LDA models, but its contribution to the long-time dynamics is also identical. The exact and LDA residues are \(\operatorname*{Res}[\widetilde c(-i\Omega)]=[1-\partial_E Y(E)|_{E=\Omega}]^{-1}\) and \(\operatorname*{Res}[\widetilde c_{\mathrm{LDA}}(-i\Omega)]=[1-\partial_E Y_{\mathrm{LDA}}(E)|_{E=\Omega}]^{-1}\), respectively. To compare the long-time rBIC contributions in the two models, we differentiate Eqs.~\eqref{eq:SM_BIC_Lamb_shift} and \eqref{eq:SM_LDA_Lamb_shift} at \(E=\Omega\) and impose the rBIC energy condition \(Y(\Omega)=\Omega\) to obtain
\begin{equation}
\begin{aligned}
\partial_E Y(E)|_{E=\Omega}&=\partial_E Y_{\mathrm{LDA}}(E)|_{E=\Omega}\\
&=\frac{1}{v^2}\sum_{m,m'}\lambda_m\lambda_{m'}[d_{mm'}^-\cos(k_ed_{mm'}^-)-d_{mm'}^+\cos(k_ed_{mm'}^+)].
\end{aligned}
\label{eq:SM_rBIC_derivative_preservation}
\end{equation}
Equation~\eqref{eq:SM_rBIC_derivative_preservation} therefore implies \(\operatorname*{Res}[\widetilde c(-i\Omega)]=\operatorname*{Res}[\widetilde c_{\mathrm{LDA}}(-i\Omega)]\). The rBIC contribution to the dynamics is thus unchanged under the LDA, whereas the oBIC contribution generally differs due to $E \neq \Omega$.

\section{Extension to multiple giant atoms} \label{S3}
\subsection{Collective dynamical equation}
The \(N\)-atom Hamiltonian is Eq.~(A1) in the End Matter. At time \(t\), we write the state of the total system as \(|\Psi_N(t)\rangle=[\sum_n c_n(t)\hat\sigma_n^\dagger+\sum_k d_k(t)\hat b_k^\dagger]|g^{\otimes N},\{0_k\}\rangle\) in the single-excitation subspace. We define the atomic probability-amplitude vector \(\boldsymbol c(t)=[c_1(t),\ldots,c_N(t)]^{\mathsf T}\) under the initial condition \(c_{1}(0)=1\) and \(c_{n\neq1}(0)=d_k(0)=0\). From the Schr\"odinger equation, we obtain the atomic probability-amplitude equation \(i\dot c_n(t)=\Omega c_n(t)+\sum_k g_{kn}^{*}d_k(t)\) and the field amplitude equation \(i\dot d_k(t)=\omega_k d_k(t)+\sum_{n'}g_{kn'}c_{n'}(t)\). Solving the field amplitude equation for \(d_k(t)\), we have \(d_k(t)=-i\sum_{n'}g_{kn'}\int_0^t d\tau\,e^{-i\omega_k(t-\tau)}c_{n'}(\tau)\). Substituting the solution for \(d_k(t)\) into the atomic amplitude equation, we obtain the matrix-form dynamical equation (A2) in the End matter
\begin{equation}
\dot{\boldsymbol c}(t)+i\Omega\boldsymbol c(t)+\int_0^t d\tau\,\boldsymbol G(t-\tau)\boldsymbol c(\tau)=\mathbf0,\label{eq:SM_collective_exact_memory_equation}
\end{equation}
where \(\boldsymbol G(t)\) is the environmental correlation matrix and its elements are
\begin{equation}
G_{nn'}(t)=\sum_k g_{kn}^{*}g_{kn'}e^{-i\omega_k t}=\int_{\omega_-}^{\omega_+}d\omega\,J_{nn'}(\omega)e^{-i\omega t},\label{eq:SM_collective_correlation}
\end{equation}
with \(J_{nn'}(\omega)\equiv[\boldsymbol J(\omega)]_{nn'}=\sum_k g_{kn}^{*}g_{kn'}\delta(\omega-\omega_k)\) being the element of the spectral density matrix \(\boldsymbol J(\omega)\). In the continuum limit, the elements \(J_{nn'}(\omega)\) can be analytically obtained as
\begin{align}
J_{nn'}(\omega)&=\frac{2}{\pi}\sum_{m=1}^{M_n}\sum_{m'=1}^{M_{n'}}\lambda_{nm}\lambda_{n'm'}\int_0^\pi dk\,\sin(kx_{nm})\sin(kx_{n'm'})\delta(\omega-\omega_k)\notag\\
&=\frac{2\rho(\omega)}{\pi}\sum_{m=1}^{M_n}\sum_{m'=1}^{M_{n'}}\lambda_{nm}\lambda_{n'm'}\sin(k_\omega x_{nm})\sin(k_\omega x_{n'm'}).\label{eq:SM_collective_exact_spectral_density}
\end{align}

\subsection{Energy spectrum with multiple giant atoms}
The eigenstate \(|\Phi_N\rangle\) of \(\hat H_N\) can be written as \(|\Phi_N\rangle=[\sum_n\alpha_n\hat\sigma_n^\dagger+\sum_k\beta_k\hat b_k^\dagger]|g^{\otimes N},\{0_k\}\rangle\) in the single-excitation subspace. From the eigenvalue equation $\hat H_N|\Phi_N\rangle=E|\Phi_N\rangle$, we obtain the coupled equations \(E\alpha_n=\Omega\alpha_n+\sum_k g_{kn}^{*}\beta_k\) and \(E\beta_k=\omega_k\beta_k+\sum_{n'}g_{kn'}\alpha_{n'}\) for the eigenstate coefficients. Solving the equation for the field coefficient \(\beta_k\), we have \(\beta_k=\sum_{n'}g_{kn'}\alpha_{n'}/(E-\omega_k)\). Plugging the solution for \(\beta_k\) into the equation for the atomic coefficient \(\alpha_n\), we obtain the matrix-form energy transcendental equation (A3) in the End matter
\begin{equation}
\boldsymbol Y(E)\boldsymbol\alpha\equiv[\Omega\mathbf I+\int_{\omega_-}^{\omega_+}d\omega\,\frac{\boldsymbol J(\omega)}{E-\omega}]\boldsymbol\alpha=E\boldsymbol\alpha,\label{eq:SM_collective_energy_equation}
\end{equation}
where \(\boldsymbol\alpha=[\alpha_1,\ldots,\alpha_N]^{\mathsf T}\) is the atomic probability-amplitude vector.
Taking the Laplace transform of Eq.~\eqref{eq:SM_collective_exact_memory_equation}, we obtain \(\widetilde{\boldsymbol c}(s)=[s\mathbf I+i\Omega\mathbf I+\widetilde{\boldsymbol G}(s)]^{-1}\boldsymbol c(0)\), where \(\widetilde{\boldsymbol G}(s)\) is the Laplace transform of \(\boldsymbol G(t)\). The possible poles of \(\widetilde{\boldsymbol c}(s)\) occur where the matrix \(s\mathbf I+i\Omega\mathbf I+\widetilde{\boldsymbol G}(s)\) is singular. Equivalently, the pole equation of \(\widetilde{\boldsymbol c}(s)\) can be written as 
\begin{equation}
\det[s\mathbf I+i\Omega\mathbf I+\widetilde{\boldsymbol G}(s)]=0.\label{eq:SM_collective_pole_equation}
\end{equation}
Setting $s=-iE$ in Eq.~\eqref{eq:SM_collective_pole_equation} recovers the energy equation in Eq.~\eqref{eq:SM_collective_energy_equation}.

\subsection{Bound-state conditions with multiple giant atoms}
We denote the integral in Eq.~\eqref{eq:SM_collective_energy_equation} by \(\boldsymbol\Delta(E)\equiv\int_{\omega_-}^{\omega_+}d\omega\,\boldsymbol J(\omega)/(E-\omega)\) and evaluate its matrix elements for \(E<\omega_-\) or \(E>\omega_+\) analogously to Eq.~\eqref{eq:SM_edge_shift_k_integral} to obtain
\begin{equation}
\Delta_{nn'}(E)=\frac{\operatorname{sgn}(E-\omega_0)}{\sqrt{(E-\omega_0)^2-4\lambda_0^2}}\sum_{m=1}^{M_n}\sum_{m'=1}^{M_{n'}}\lambda_{nm}\lambda_{n'm'}[\zeta_E^{|x_{nm}-x_{n'm'}|}-\zeta_E^{x_{nm}+x_{n'm'}}].\label{eq:SM_collective_shift_integral}
\end{equation}
The energy transcendental equation~\eqref{eq:SM_collective_energy_equation} then reads
\begin{equation}
\boldsymbol Y(E)\boldsymbol\alpha\equiv[\Omega\mathbf I+\boldsymbol\Delta(E)]\boldsymbol\alpha=E\boldsymbol\alpha.\label{eq:SM_collective_BOC_energy_equation}
\end{equation}
The real roots with \(E<\omega_-\) or \(E>\omega_+\) determine the BOC energies. Introducing the band-edge matrices by \(\boldsymbol\Delta^-\equiv-\lim_{E\to\omega_-^-}\boldsymbol\Delta(E)\) and \(\boldsymbol\Delta^+\equiv\lim_{E\to\omega_+^+}\boldsymbol\Delta(E)\) and taking the limits as in Eqs.~\eqref{eq:SM_lower_edge_shift} and \eqref{eq:SM_upper_edge_shift}, we obtain 
\begin{align}
[\boldsymbol\Delta^-]_{nn'}\equiv\Delta_{nn'}^-
&=\frac{1}{\lambda_0}\sum_{m=1}^{M_n}\sum_{m'=1}^{M_{n'}}\lambda_{nm}\lambda_{n'm'}\min(x_{nm},x_{n'm'}),\label{eq:SM_collective_lower_edge_matrix}\\
[\boldsymbol\Delta^+]_{nn'}\equiv\Delta_{nn'}^+
&=\frac{1}{\lambda_0}\sum_{m=1}^{M_n}\sum_{m'=1}^{M_{n'}}(-1)^{x_{nm}+x_{n'm'}}\lambda_{nm}\lambda_{n'm'}\min(x_{nm},x_{n'm'}).
\label{eq:SM_collective_upper_edge_matrix}
\end{align}
According to Eq.~\eqref{eq:SM_collective_BOC_energy_equation}, a BOC exists when the matrix
\(E\mathbf I-\boldsymbol Y(E)\) has a zero eigenvalue. The eigenvalues of the matrix \(E\mathbf I-\boldsymbol Y(E)\) increase strictly with \(E\) in the regimes \(E<\omega_-\) and \(E>\omega_+\). As \(E\) increases from \(-\infty\) to \(\omega_-\), the eigenvalues of the matrix \(E\mathbf I-\boldsymbol Y(E)\) rise from negative values to those of \(\boldsymbol\Delta^--(\Omega-\omega_-)\mathbf I\). The positive eigenvalue of \(\boldsymbol\Delta^--(\Omega-\omega_-)\mathbf I\) therefore implies the existence of a BOC with \(E<\omega_-\). As \(E\) increases from \(\omega_+\) to \(+\infty\), the eigenvalues of the matrix \(E\mathbf I-\boldsymbol Y(E)\) rise from those of \((\omega_+-\Omega)\mathbf I-\boldsymbol\Delta^+\) to positive values. The BOC with \(E>\omega_+\) exists when \((\omega_+-\Omega)\mathbf I-\boldsymbol\Delta^+\) has a negative eigenvalue. The BOC conditions can be written as
\begin{align}
\Delta_j^-&>\Omega-\omega_-,\qquad E_j^{\mathrm{boc}}<\omega_-,
\label{eq:SM_collective_lower_BOC_condition}\\
\Delta_j^+&>\omega_+-\Omega,\qquad E_j^{\mathrm{boc}}>\omega_+,
\label{eq:SM_collective_upper_BOC_condition}
\end{align}
where \(\Delta_j^\pm\) represents the eigenvalues of \(\boldsymbol\Delta^\pm\).

For the BIC with energy \(\omega_-<E_j^{\mathrm{bic}}<\omega_+\), we obtain BIC conditions analogous to those in Eqs.~\eqref{eq:SM_BIC_dark_condition} and \eqref{eq:SM_BIC_energy_condition} for the single giant atom case
\begin{align}
\boldsymbol J(E_j^{\mathrm{bic}})\boldsymbol\alpha_j&=\mathbf 0,
\label{eq:SM_collective_BIC_matrix_darkness}\\
[(E_j^{\mathrm{bic}}-\Omega)\mathbf I-\boldsymbol\Delta(E_j^{\mathrm{bic}})]\boldsymbol\alpha_j&=\mathbf 0,\label{eq:SM_collective_BIC_energy_condition}
\end{align}
where \(\boldsymbol\alpha_j\) is the atomic probability-amplitude vector of the BIC. Plugging Eq.~\eqref{eq:SM_collective_exact_spectral_density} into Eq.~\eqref{eq:SM_collective_BIC_matrix_darkness}, we obtain the explicit form of Eq.~\eqref{eq:SM_collective_BIC_matrix_darkness}
\begin{equation}
\sum_{n=1}^{N}\sum_{m=1}^{M_n}\lambda_{nm}\sin(k_jx_{nm})\alpha_{j\,n}=0.
\label{eq:SM_collective_BIC_darkness}
\end{equation}
To write the BIC condition in Eq.~\eqref{eq:SM_collective_BIC_energy_condition} in explicit form, we evaluate the matrix elements of \(\boldsymbol\Delta(E_j^{\mathrm{bic}})\) as in Eq.~\eqref{eq:SM_BIC_Lamb_shift} and obtain
\begin{equation}
\Delta_{nn'}(E_j^{\mathrm{bic}})=\frac{1}{v_j}\sum_{m=1}^{M_n}\sum_{m'=1}^{M_{n'}}\lambda_{nm}\lambda_{n'm'}[\sin(k_jd_{nm\,n'm'}^-)-\sin(k_jd_{nm\,n'm'}^+)].
\label{eq:SM_collective_BIC_shift}
\end{equation}

\subsection{Long-time dynamics for multiple giant atoms}
Following the contour deformation in Eq.~\eqref{eq:SM_pole_cut_decomposition}, we introduce $\boldsymbol A(s)=s\mathbf I+i\Omega\mathbf I+\widetilde{\boldsymbol G}(s)$ and write the inverse Laplace transform of $\widetilde{\boldsymbol c}(s)$ as
\begin{equation}
\boldsymbol c(t)=\sum_{\nu=\mathrm{boc,bic}}\sum_j\operatorname*{Res}_{s=-iE_j^\nu}[e^{st}\boldsymbol A^{-1}(s)]\boldsymbol c(0)+\frac{1}{2\pi i}\int_{\Gamma_{\mathrm{cut}}}ds\,e^{st}\boldsymbol A^{-1}(s)\boldsymbol c(0).
\label{eq:SM_collective_contour}
\end{equation}
Denoting the residue matrix as $\boldsymbol R_j^\nu\equiv\operatorname*{Res}_{s=-iE_j^\nu}\boldsymbol A^{-1}(s)$, the contribution from each bound-state \(j\) is then $\boldsymbol R_j^\nu\boldsymbol c(0)e^{-iE_j^\nu t}$. As in the single-atom case, the branch-cut contribution vanishes in the long-time limit and we obtain Eq.~(A7) in the End matter
\begin{equation}
\boldsymbol c(\infty)
=\sum_{\nu=\mathrm{boc,bic}}\sum_j
\boldsymbol R_j^\nu\boldsymbol c(0)e^{-iE_j^\nu t}.
\label{eq:SM_collective_long_time_solution}
\end{equation}
To evaluate the residue matrix $\boldsymbol R_j^\nu$, we set $s_j=-iE_j^\nu$. The BOC condition in Eq.~\eqref{eq:SM_collective_BOC_energy_equation} or the BIC conditions in Eqs.~\eqref{eq:SM_collective_BIC_matrix_darkness} and \eqref{eq:SM_collective_BIC_energy_condition} give $\boldsymbol A(s_j)\boldsymbol\alpha_j^\nu=\mathbf 0$, where $\boldsymbol\alpha_j^\nu$ is the atomic probability-amplitude vector of the bound state. The corresponding BOC or BIC conditions, together with the Hermiticity of $\boldsymbol J(E_j^\nu)$ and $\boldsymbol\Delta(E_j^\nu)$, also imply $(\boldsymbol\alpha_j^\nu)^\dagger\boldsymbol A(s_j)=\mathbf 0$. Expanding $\boldsymbol A(s)$ and its inverse around $s_j$, we write
\begin{align}
\boldsymbol A(s_j+\epsilon)&=\boldsymbol A_0+\epsilon\boldsymbol A_1+\mathcal O(\epsilon^2),\notag\\
\boldsymbol A^{-1}(s_j+\epsilon)&=\frac{\boldsymbol B_{-1}}{\epsilon}+\boldsymbol B_0+\mathcal O(\epsilon),
\label{eq:SM_collective_residue_expansions}
\end{align}
where $\boldsymbol A_0=\boldsymbol A(s_j)$, $\boldsymbol A_1=\boldsymbol A'(s_j)$, and $\boldsymbol B_{-1}=\boldsymbol R_j^\nu$. Comparing the coefficients of $\epsilon^{-1}$
and $\epsilon^0$ in $\boldsymbol A(s)\boldsymbol A^{-1}(s)=\boldsymbol A^{-1}(s)\boldsymbol A(s)=\mathbf I$, we obtain
\begin{equation}
\boldsymbol A_0\boldsymbol R_j^\nu=\mathbf 0,\qquad
\boldsymbol R_j^\nu\boldsymbol A_0=\mathbf 0,\qquad
\boldsymbol A_0\boldsymbol B_0+
\boldsymbol A_1\boldsymbol R_j^\nu=\mathbf I.
\label{eq:SM_collective_residue_forward_orders}
\end{equation}
For the sake of simplicity, we consider a nondegenerate bound state, thus the left and right null spaces of $\boldsymbol A_0$ are spanned by $(\boldsymbol\alpha_j^\nu)^\dagger$ and $\boldsymbol\alpha_j^\nu$, respectively.
The first two relations in Eq.~\eqref{eq:SM_collective_residue_forward_orders} therefore give
$\boldsymbol R_j^\nu=r_j^\nu\boldsymbol\alpha_j^\nu(\boldsymbol\alpha_j^\nu)^\dagger$, where $r_j^\nu$ is a scalar coefficient. Multiplying the third relation in Eq.~\eqref{eq:SM_collective_residue_forward_orders} from the left by $(\boldsymbol\alpha_j^\nu)^\dagger$ and substituting $\boldsymbol R_j^\nu=r_j^\nu\boldsymbol\alpha_j^\nu(\boldsymbol\alpha_j^\nu)^\dagger$, we obtain $r_j^\nu=[(\boldsymbol\alpha_j^\nu)^\dagger\boldsymbol A_1\boldsymbol\alpha_j^\nu]^{-1}$. With $\boldsymbol A_1=\boldsymbol A'(s_j)$ and the self-energy matrix $\boldsymbol\Delta(E)$ defined in Eq.~\eqref{eq:SM_collective_BOC_energy_equation}, we find
\begin{equation}
(\boldsymbol\alpha_j^\nu)^\dagger\boldsymbol A'(s_j)\boldsymbol\alpha_j^\nu=(\boldsymbol\alpha_j^\nu)^\dagger[\mathbf I-\partial_E\boldsymbol\Delta(E)|_{E=E_j^\nu}]\boldsymbol\alpha_j^\nu.
\label{eq:SM_collective_residue_projected_derivative}
\end{equation}
The residue matrix is therefore calculated as
\begin{equation}
\boldsymbol R_j^\nu=\frac{\boldsymbol\alpha_j^\nu(\boldsymbol\alpha_j^\nu)^\dagger}{(\boldsymbol\alpha_j^\nu)^\dagger[\mathbf I-\partial_E\boldsymbol\Delta(E)|_{E=E_j^\nu}]\boldsymbol\alpha_j^\nu}.
\label{eq:SM_collective_residue}
\end{equation}

\subsection{Collective LDA dynamical equation}
Applying the same dispersion linearization as in Eq.~\eqref{eq:SM_LDA_spectral_density} to the collective spectral-density matrix \(J_{nn'}(\omega)\), we obtain its LDA version
\begin{align}
J_{{\mathrm{LDA}}\,nn'}(\omega)&=\sum_k g_{kn}^{*}g_{kn'}\delta(\omega-\omega_k^{\mathrm{LDA}})\notag\\
&=\frac{2}{\pi v}\sum_{m=1}^{M_n}\sum_{m'=1}^{M_{n'}}\lambda_{nm}\lambda_{n'm'}\sin(k_\omega^{\mathrm{LDA}}x_{nm})\sin(k_\omega^{\mathrm{LDA}}x_{n'm'}).\label{eq:SM_collective_LDA_spectral_density}
\end{align}
The environmental correlation matrix under the LDA is $\boldsymbol G_{\mathrm{LDA}}(t)=\int_{-\infty}^{\infty}d\omega\,\boldsymbol J_{\mathrm{LDA}}(\omega)e^{-i\omega t}$. Substituting this correlation matrix \(\boldsymbol G_{\mathrm{LDA}}(t)\) into Eq.~\eqref{eq:SM_collective_exact_memory_equation} and evaluating the time convolution as in the single-atom derivation of Eq.~\eqref{eq:SM_LDA_delay_equation}, we obtain the LDA dynamical equation (A8) in the End Matter
\begin{align}
\dot c_{{\mathrm{LDA}}\,n}(t)&=-i\Omega c_{{\mathrm{LDA}}\,n}(t)\notag\\
&\quad-\sum_{n'=1}^{N}\sum_{m=1}^{M_n}\sum_{m'=1}^{M_{n'}}\Gamma_{nm\,n'm'}^-c_{{\mathrm{LDA}}\,n'}(t-\tau_{nm\,n'm'}^-)\Theta(t-\tau_{nm\,n'm'}^-)\notag\\
&\quad+\sum_{n'=1}^{N}\sum_{m=1}^{M_n}\sum_{m'=1}^{M_{n'}}\Gamma_{nm\,n'm'}^+c_{{\mathrm{LDA}}\,n'}(t-\tau_{nm\,n'm'}^+)\Theta(t-\tau_{nm\,n'm'}^+).\label{eq:SM_collective_LDA_delay_equation}
\end{align}

\subsection{Energy spectrum of the LDA with multiple giant atoms}
Collecting the elementwise equations in Eq.~\eqref{eq:SM_collective_LDA_delay_equation} into the vector $\boldsymbol c_{\mathrm{LDA}}(t)=(c_{\mathrm{LDA}\,1}(t),\ldots,c_{\mathrm{LDA}\,N}(t))^{\mathsf T}$ and taking the Laplace transform, we obtain $\widetilde{\boldsymbol c}_{\mathrm{LDA}}(s)=\boldsymbol A_{\mathrm{LDA}}^{-1}(s)\boldsymbol c(0)$, where $\boldsymbol A_{\mathrm{LDA}}(s)=s\mathbf I+i\Omega\mathbf I+\widetilde{\boldsymbol G}_{\mathrm{LDA}}(s)$ and $\widetilde{\boldsymbol G}_{\mathrm{LDA}}(s)=\int_{-\infty}^{\infty}d\omega\,\boldsymbol J_{\mathrm{LDA}}(\omega)/(s+i\omega)$ denotes the Laplace transform of \(\boldsymbol G_{\mathrm{LDA}}(t)\). Setting $s=-iE$ in the singularity condition $\det A_{\rm LDA}(s)=0$ and imposing the LDA BIC conditions (Eqs.~\eqref{eq:SM_collective_LDA_BIC_matrix_darkness} and \eqref{eq:SM_collective_LDA_BIC_energy_condition}), we obtain
\begin{equation}
\boldsymbol Y_{\mathrm{LDA}}(\bar E_j^{\mathrm{bic}})\bar{\boldsymbol\alpha}_j=\bar E_j^{\mathrm{bic}}\bar{\boldsymbol\alpha}_j.
\label{eq:SM_collective_LDA_energy_equation_at_BIC}
\end{equation}
Replacing $\boldsymbol J(\omega)$ with $\boldsymbol J_{\mathrm{LDA}}(\omega)$ and extending the integration limits in Eq.~\eqref{eq:SM_collective_energy_equation} to $-\infty$ and $+\infty$, we obtain the matrix-form LDA energy transcendental equation (A9) in the End Matter
\begin{equation}
\boldsymbol Y_{\mathrm{LDA}}(E)\bar{\boldsymbol\alpha}\equiv[\Omega\mathbf I+\boldsymbol\Delta_{\mathrm{LDA}}(E)]\bar{\boldsymbol\alpha}=E\bar{\boldsymbol\alpha},
\label{eq:SM_collective_LDA_energy_equation}
\end{equation}
where $\bar{\boldsymbol\alpha}=(\bar\alpha_1,\ldots,\bar\alpha_N)^{\mathsf T}$ is the atomic probability-amplitude vector of an LDA eigenstate with energy $E$, and $\boldsymbol\Delta_{\mathrm{LDA}}(E)=\int_{-\infty}^{\infty}d\omega\,\boldsymbol J_{\mathrm{LDA}}(\omega)/(E-\omega)$. It is evident that Eq.~\eqref{eq:SM_collective_LDA_energy_equation_at_BIC} is precisely Eq.~\eqref{eq:SM_collective_LDA_energy_equation} by setting $E=\bar E_j^{\mathrm{bic}}$. The matrix elements of $\boldsymbol\Delta_{\mathrm{LDA}}(E)$ can be evaluated as in Eq.~\eqref{eq:SM_LDA_Lamb_shift}
\begin{equation}
\Delta_{\mathrm{LDA}\,nn'}(E)=\frac{1}{v}\sum_{m=1}^{M_n}\sum_{m'=1}^{M_{n'}}\lambda_{nm}\lambda_{n'm'}[\sin(k_E^{\mathrm{LDA}}d_{nm\,n'm'}^-)-\sin(k_E^{\mathrm{LDA}}d_{nm\,n'm'}^+)].
\label{eq:SM_collective_LDA_shift}
\end{equation}

\subsection{Collective LDA bound-state conditions}
As in the single-giant-atom case, all bound states under the LDA are BICs because the linearized dispersion has no band edges. 
Following the derivation of the single-giant-atom BIC conditions in Eqs.~\eqref{eq:SM_LDA_BIC_dark_condition} and \eqref{eq:SM_LDA_BIC_energy_condition}, we obtain the conditions for forming the $j$th collective LDA BIC with energy $\bar E_j^{\mathrm{bic}}$ as
\begin{align}
\boldsymbol J_{\mathrm{LDA}}(\bar E_j^{\mathrm{bic}})\bar{\boldsymbol\alpha}_j&=\mathbf 0, \label{eq:SM_collective_LDA_BIC_matrix_darkness}\\
[\bar E_j^{\mathrm{bic}}\mathbf I-\boldsymbol Y_{\mathrm{LDA}}(\bar E_j^{\mathrm{bic}})]\bar{\boldsymbol\alpha}_j&=\mathbf 0. \label{eq:SM_collective_LDA_BIC_energy_condition}
\end{align}
The LDA condition in Eq.~\eqref{eq:SM_collective_LDA_BIC_matrix_darkness} has the same form as the exact condition in Eq.~\eqref{eq:SM_collective_BIC_darkness}, with $k_j$ and $\alpha_{j\,n}$ replaced by $\bar k_j$ and $\bar\alpha_{j\,n}$, respectively. As in the single-giant-atom analysis, the exact and LDA BIC conditions coincide for rBICs but generally differ for oBICs.

\subsection{Collective LDA long-time dynamics}
As in the single-atom case, only the LDA BIC contributions remain in the long-time limit
\begin{equation}
\boldsymbol c_{\mathrm{LDA}}(\infty)=\sum_j\bar{\boldsymbol R}_j\boldsymbol c(0)e^{-i\bar E_j^{\mathrm{bic}}t}.
\label{eq:SM_collective_LDA_long_time_solution}
\end{equation}
Following the same residue calculation in Eq.~\eqref{eq:SM_collective_residue}, the LDA residue matrix is
\begin{equation}
\bar{\boldsymbol R}_j=\frac{\bar{\boldsymbol\alpha}_j\bar{\boldsymbol\alpha}_j^\dagger}{\bar{\boldsymbol\alpha}_j^\dagger\left[\mathbf I-\left.\partial_E\boldsymbol\Delta_{\mathrm{LDA}}(E)\right|_{E=\bar E_j^{\mathrm{bic}}}\right]\bar{\boldsymbol\alpha}_j}.
\label{eq:SM_collective_LDA_residue}
\end{equation}
To compare the exact and LDA residue matrices for rBICs, we differentiate $\boldsymbol\Delta(E)$ and $\boldsymbol\Delta_{\mathrm{LDA}}(E)$ and evaluate their difference at $E=\Omega$, obtaining $\left.\partial_E\boldsymbol\Delta(E)\right|_{E=\Omega}-\left.\partial_E\boldsymbol\Delta_{\mathrm{LDA}}(E)\right|_{E=\Omega}=(\Omega-\omega_0)\boldsymbol\Delta(\Omega)/v^2$. 
For an rBIC, the energy condition in Eq.~\eqref{eq:SM_collective_BIC_energy_condition} requires $\boldsymbol\Delta(\Omega)\boldsymbol\alpha_j=\mathbf 0$. Together with the derivative relation above, this condition ensures that the exact and LDA derivative matrices act identically on $\boldsymbol\alpha_j$, namely, $\left.\partial_E\boldsymbol\Delta(E)\right|_{E=\Omega}\boldsymbol\alpha_j=\left.\partial_E\boldsymbol\Delta_{\mathrm{LDA}}(E)\right|_{E=\Omega}\boldsymbol\alpha_j$.
Since the exact and LDA rBIC conditions coincide, we can choose $\bar{\boldsymbol\alpha}_j=\boldsymbol\alpha_j$. Consequently, the projected derivatives entering the exact and LDA residue matrices are identical:
\begin{equation}
\begin{aligned}
\boldsymbol\alpha_j^\dagger\left[\left.\partial_E\boldsymbol\Delta(E)\right|_{E=\Omega}\right]\boldsymbol\alpha_j&=\boldsymbol\alpha_j^\dagger\left[\left.\partial_E\boldsymbol\Delta_{\mathrm{LDA}}(E)\right|_{E=\Omega}\right]\boldsymbol\alpha_j\\
&=\frac{1}{v^2}\sum_{n,n'=1}^{N}\sum_{m=1}^{M_n}\sum_{m'=1}^{M_{n'}}\alpha_{j\,n}^{*}\lambda_{nm}\lambda_{n'm'}\alpha_{j\,n'}\\
&\quad\times\Bigl[d_{nm\,n'm'}^-\cos(k_ed_{nm\,n'm'}^-)-d_{nm\,n'm'}^+\cos(k_ed_{nm\,n'm'}^+)\Bigr].
\end{aligned}
\label{eq:SM_collective_rBIC_derivative_preservation}
\end{equation}
Since the exact and LDA rBIC conditions coincide and the corresponding residue matrices are identical, comparing the long-time solutions in Eqs.~\eqref{eq:SM_collective_long_time_solution} and \eqref{eq:SM_collective_LDA_long_time_solution} shows that the rBIC contributions to the long-time dynamics are exactly preserved under the LDA, whereas the oBIC contributions generally differ.

\section{Extension to infinite waveguide}
\subsection{Exact dynamical equation}
We now extend the results obtained in Sec.~\ref{S3} to an infinite-waveguide configuration. Starting from an $L$-resonator array with periodic boundary conditions, we introduce the Fourier transform $\hat b_x=L^{-1/2}\sum_k e^{ikx}\hat b_k$ and take the limit $L\to\infty$. The resulting momentum-space Hamiltonian reads
\begin{equation}
\hat H_N^{\mathrm{inf}}
=\Omega\sum_{n=1}^{N}\hat\sigma_n^\dagger\hat\sigma_n
+\sum_k\omega_k\hat b_k^\dagger\hat b_k
+\sum_{k,n}(g_{kn}^{\mathrm{inf}}\hat\sigma_n\hat b_k^\dagger+\mathrm{H.c.}),
\label{eq:SM_inf_momentum_Hamiltonian}
\end{equation}
in which the dispersion relation remains unchanged as $\omega_k=\omega_0-2\lambda_0\cos k$ and the coupling coefficients take the form $g_{kn}^{\mathrm{inf}}=L^{-1/2}\sum_{m=1}^{M_n}\lambda_{nm}e^{-ikx_{nm}}$. 
Following the same procedure as in the derivation of Eq.~\eqref{eq:SM_collective_exact_memory_equation}, we eliminate the photonic amplitudes and obtain the matrix-form dynamical equation
\begin{equation}
\dot{\boldsymbol c}(t)+i\Omega\boldsymbol c(t)+\int_0^t d\tau\,\boldsymbol G^{\mathrm{inf}}(t-\tau)\boldsymbol c(\tau)=\mathbf0,
\label{eq:SM_inf_dynamical_equation}
\end{equation}
where $\boldsymbol G^{\mathrm{inf}}(t)=\int_{\omega_-}^{\omega_+}d\omega\,\boldsymbol J^{\mathrm{inf}}(\omega)e^{-i\omega t}$.
For the infinite waveguide, the elements of the spectral-density matrix in the continuum limit $\sum_k\to L\int_{-\pi}^{\pi}dk/(2\pi)$ are derived as
\begin{align}
J_{nn'}^{\mathrm{inf}}(\omega)&=\sum_k(g_{kn}^{\mathrm{inf}})^{*}g_{kn'}^{\mathrm{inf}}\delta(\omega-\omega_k)\notag\\
&=\frac{1}{2\pi}\sum_{m=1}^{M_n}\sum_{m'=1}^{M_{n'}}\lambda_{nm}\lambda_{n'm'}\int_{-\pi}^{\pi}dk\,e^{ik(x_{nm}-x_{n'm'})}\delta(\omega-\omega_k)\notag\\
&=\frac{\rho(\omega)}{2\pi}\sum_{m=1}^{M_n}\sum_{m'=1}^{M_{n'}}\lambda_{nm}\lambda_{n'm'}[e^{ik_\omega(x_{nm}-x_{n'm'})}+e^{-ik_\omega(x_{nm}-x_{n'm'})}]\notag\\
&=\frac{\rho(\omega)}{\pi}\sum_{m=1}^{M_n}\sum_{m'=1}^{M_{n'}}\lambda_{nm}\lambda_{n'm'}\cos[k_\omega(x_{nm}-x_{n'm'})].
\label{eq:SM_inf_spectral_density}
\end{align}

\subsection{Exact energy spectrum}
Following the same procedure as in the derivation of Eq.~\eqref{eq:SM_collective_energy_equation}, we obtain the exact energy transcendental equation
\begin{equation}
\boldsymbol Y^{\mathrm{inf}}(E)\boldsymbol\alpha
\equiv [\Omega\mathbf I+\boldsymbol\Delta^{\mathrm{inf}}(E)]\boldsymbol\alpha
=E\boldsymbol\alpha,
\label{eq:SM_inf_energy_equation}
\end{equation}
where $\boldsymbol\Delta^{\mathrm{inf}}(E)=\int_{\omega_-}^{\omega_+}d\omega\,\boldsymbol J^{\mathrm{inf}}(\omega)/(E-\omega)$. By means of the spectral-density matrix in Eq.~\eqref{eq:SM_inf_spectral_density}, the analytical expression of the matrix elements of $\boldsymbol\Delta^{\mathrm{inf}}(E)$ for energies outside the band \(E>\omega_+\) or \(E<\omega_-\) is evaluated as
\begin{align}
\Delta_{nn'}^{\mathrm{inf}}(E)&=\frac{1}{\pi}\sum_{m=1}^{M_n}\sum_{m'=1}^{M_{n'}}\lambda_{nm}\lambda_{n'm'}\int_0^\pi dk\,\frac{\cos[k(x_{nm}-x_{n'm'})]}{E-\omega_0+2\lambda_0\cos k}\notag\\
&=\frac{\operatorname{sgn}(E-\omega_0)}{\sqrt{(E-\omega_0)^2-4\lambda_0^2}}\sum_{m=1}^{M_n}\sum_{m'=1}^{M_{n'}}\lambda_{nm}\lambda_{n'm'}\zeta_E^{|x_{nm}-x_{n'm'}|}.
\label{eq:SM_inf_shift_integral}
\end{align}
For a single atom coupled to the waveguide at one site, this expression recovers the energy shift obtained in Ref.~\cite{Calajo2016}. For in-band energies $\omega_-<E<\omega_+$, the expression of the matrix elements of $\boldsymbol\Delta^{\mathrm{inf}}(E)$ is calculated as
\begin{align}
\Delta_{nn'}^{\mathrm{inf}}(E)&=\frac{1}{2\pi\lambda_0}\sum_{m=1}^{M_n}\sum_{m'=1}^{M_{n'}}\lambda_{nm}\lambda_{n'm'}\mathcal P\int_0^\pi dk\,\frac{\cos[k(x_{nm}-x_{n'm'})]}{\cos k-\cos k_E}\notag\\
&=\frac{1}{v_E}\sum_{m=1}^{M_n}\sum_{m'=1}^{M_{n'}}\lambda_{nm}\lambda_{n'm'}\sin(k_E|x_{nm}-x_{n'm'}|).
\label{eq:SM_inf_inband_shift}
\end{align} 
The pole equation for $\widetilde{\boldsymbol c}(s)$ obtained from Eq.~\eqref{eq:SM_inf_dynamical_equation} has the same form as Eq.~\eqref{eq:SM_collective_pole_equation}, with $\widetilde{\boldsymbol G}(s)$ replaced by $\widetilde{\boldsymbol G}^{\mathrm{inf}}(s)$. Setting $s=-iE$ then recovers the energy transcendental equation in Eq.~\eqref{eq:SM_inf_energy_equation}.

\subsection{Exact bound-state conditions}
The same eigenvalue analysis as in Eqs.~\eqref{eq:SM_collective_lower_BOC_condition} and \eqref{eq:SM_collective_upper_BOC_condition} leads to the BOC conditions
\begin{align}
\Delta_{\mathrm{inf}\,j}^-&>\Omega-\omega_-,\qquad E_j^{\mathrm{boc}}<\omega_-,\label{eq:SM_inf_lower_BOC_condition}\\
\Delta_{\mathrm{inf}\,j}^+&>\omega_+-\Omega,\qquad E_j^{\mathrm{boc}}>\omega_+,\label{eq:SM_inf_upper_BOC_condition}
\end{align}
where \(\Delta_{\mathrm{inf}\,j}^-\) and \(\Delta_{\mathrm{inf}\,j}^+\) are the limits of the \(j\)th eigenvalues of \(-\boldsymbol\Delta_{\mathrm{inf}}(E)\) as \(E\to\omega_-^-\) and \(\boldsymbol\Delta_{\mathrm{inf}}(E)\) as \(E\to\omega_+^+\), respectively. As in the semi-infinite waveguide, the numbers of BOCs below and above the band are obtained by counting the eigenvalues satisfying Eqs.~\eqref{eq:SM_inf_lower_BOC_condition} and \eqref{eq:SM_inf_upper_BOC_condition}, respectively. However, the limits of band-edge eigenvalue $\Delta_{\mathrm{inf}\,j}^{\pm}$ can diverge, unlike the finite limits $\Delta_j^{\pm}$ in the semi-infinite waveguide. According to Eq.~\eqref{eq:SM_inf_shift_integral}, one lower-edge eigenvalue diverges to \(+\infty\) if \(\sum_m\lambda_{nm}\ne0\) for at least one atom. One upper-edge eigenvalue diverges to \(+\infty\) if \(\sum_m(-1)^{x_{nm}}\lambda_{nm}\ne0\) for at least one atom. Each divergence ensures the existence of a BOC below or above the band, respectively. The remaining eigenvalue limits are finite and must satisfy Eqs.~\eqref{eq:SM_inf_lower_BOC_condition} and \eqref{eq:SM_inf_upper_BOC_condition} for additional BOCs to exist. With a single atom coupled at one site, the BOC exists on each side of the band for any nonzero coupling, in agreement with Ref.~\cite{SanchezBurillo2017}.

As in the semi-infinite waveguide, the BIC with energy \(E_j^{\mathrm{bic}}\in(\omega_-,\omega_+)\) satisfies 
\begin{align}
\boldsymbol J_{\mathrm{inf}}(E_j^{\mathrm{bic}})\boldsymbol\alpha_j&=\mathbf0,\label{eq:SM_inf_BIC_matrix_darkness}\\
[(E_j^{\mathrm{bic}}-\Omega)\mathbf I-\boldsymbol\Delta_{\mathrm{inf}}(E_j^{\mathrm{bic}})]\boldsymbol\alpha_j&=\mathbf0.\label{eq:SM_inf_BIC_energy_condition}
\end{align}
For the semi-infinite waveguide, Eq.~\eqref{eq:SM_collective_BIC_darkness} requires $\sum_{n,m}\lambda_{nm}\sin(k_jx_{nm})\alpha_{j\,n}=0$. In contrast, Eq.~\eqref{eq:SM_inf_spectral_density} contains contributions from both $k_j$ and $-k_j$, which must vanish independently. Therefore, Eq.~\eqref{eq:SM_inf_BIC_matrix_darkness} can be written as
\begin{equation}
\sum_{n=1}^{N}\sum_{m=1}^{M_n}\lambda_{nm}e^{\pm ik_jx_{nm}}\alpha_{j\,n}=0.
\label{eq:SM_inf_BIC_darkness}
\end{equation}
For two identical point-like atoms coupled to distinct waveguide sites with equal nonzero coupling strengths, Eqs.~\eqref{eq:SM_inf_BIC_energy_condition} and \eqref{eq:SM_inf_BIC_darkness} imply $E_j^{\mathrm{bic}}=\Omega$ and $\sin[k_j(x_{21}-x_{11})]=0$, recovering the two-atom BIC conditions in Ref.~\cite{Kocabas2016}.

\subsection{Exact long-time dynamics}
The same contour deformation as in the semi-infinite waveguide shows that the continuous-spectrum contribution vanishes at long times. The long-time dynamics is therefore determined by the bound states
\begin{equation}
\boldsymbol c(\infty)=\sum_{\nu=\mathrm{boc,bic}}\sum_j
\boldsymbol R_j^{\mathrm{inf}\,\nu}\boldsymbol c(0)e^{-iE_j^\nu t},
\label{eq:SM_inf_long_time}
\end{equation}
where $\boldsymbol R_j^{\mathrm{inf}\,\nu}$ is the residue matrix at $s=-iE_j^\nu$ and follows from Eq.~\eqref{eq:SM_collective_residue} with $\boldsymbol\Delta(E)$ replaced by $\boldsymbol\Delta^{\mathrm{inf}}(E)$.

\subsection{LDA dynamical equation}
For the infinite waveguide, we linearize the dispersion around the two resonant wave numbers $k_e$ and $-k_e$
\begin{equation}
\omega_k^{\mathrm{LDA}}=
\begin{cases}
\Omega+v(k-k_e), & k\simeq k_e,\\
\Omega-v(k+k_e), & k\simeq-k_e.
\end{cases}
\label{eq:SM_inf_LDA_dispersion}
\end{equation}
Applying the same dispersion linearization as in Eq.~\eqref{eq:SM_collective_LDA_spectral_density} to the infinite-waveguide spectral-density matrix in Eq.~\eqref{eq:SM_inf_spectral_density}, we obtain
\begin{equation}
J_{nn'}^{\mathrm{inf,LDA}}(\omega)=\frac{1}{\pi v}\sum_{m=1}^{M_n}\sum_{m'=1}^{M_{n'}}\lambda_{nm}\lambda_{n'm'}\cos[k_\omega^{\mathrm{LDA}}(x_{nm}-x_{n'm'})].
\label{eq:SM_inf_LDA_spectral_density}
\end{equation}
The LDA environmental correlation matrix is $\boldsymbol G^{\mathrm{inf,LDA}}(t)=\int_{-\infty}^{\infty}d\omega\,\boldsymbol J^{\mathrm{inf,LDA}}(\omega)e^{-i\omega t}$. Substituting \(\boldsymbol G^{\mathrm{inf,LDA}}(t)\) into Eq.~\eqref{eq:SM_inf_dynamical_equation} and evaluating the time convolution as in the derivation of Eq.~\eqref{eq:SM_LDA_delay_equation}, we obtain the collective LDA dynamical equation
\begin{equation}
\dot c_{\mathrm{LDA}\,n}(t)=-i\Omega c_{\mathrm{LDA}\,n}(t)-\sum_{n'=1}^{N}\sum_{m=1}^{M_n}\sum_{m'=1}^{M_{n'}}\Gamma_{nm\,n'm'}^-c_{\mathrm{LDA}\,n'}(t-\tau_{nm\,n'm'}^-)\Theta(t-\tau_{nm\,n'm'}^-).
\label{eq:SM_inf_LDA_delay_equation}
\end{equation}
Only the direct-transfer terms remain in this equation compared with Eq.~\eqref{eq:SM_collective_LDA_delay_equation}. The $\Gamma_{nm\,n'm'}^+$ terms in Eq.~\eqref{eq:SM_collective_LDA_delay_equation} describe mirror-reflected paths and are absent because the infinite waveguide has no reflecting boundary.

\subsection{LDA energy spectrum}
Following the same procedure as in Eq.~\eqref{eq:SM_collective_LDA_energy_equation}, we obtain the LDA energy transcendental equation for the infinite waveguide
\begin{equation}
\boldsymbol Y^{\mathrm{inf,LDA}}(E)\bar{\boldsymbol\alpha}
\equiv[\Omega\mathbf I+\boldsymbol\Delta^{\mathrm{inf,LDA}}(E)]\bar{\boldsymbol\alpha}
=E\bar{\boldsymbol\alpha},
\label{eq:SM_inf_LDA_energy_equation}
\end{equation}
where $\boldsymbol\Delta^{\mathrm{inf,LDA}}(E)=\int_{-\infty}^{\infty}d\omega\,\boldsymbol J^{\mathrm{inf,LDA}}(\omega)/(E-\omega)$. Using the spectral-density matrix in Eq.~\eqref{eq:SM_inf_LDA_spectral_density} and evaluating \(\boldsymbol\Delta^{\mathrm{inf,LDA}}(E)\) as in Eq.~\eqref{eq:SM_LDA_Lamb_shift}, we obtain
\begin{equation}
\Delta_{nn'}^{\mathrm{inf,LDA}}(E)
=\frac{1}{v}\sum_{m=1}^{M_n}\sum_{m'=1}^{M_{n'}}\lambda_{nm}\lambda_{n'm'}\sin(k_E^{\mathrm{LDA}}d_{nm\,n'm'}^-).
\label{eq:SM_inf_LDA_shift}
\end{equation}
Taking the Laplace transform of Eq.~\eqref{eq:SM_inf_LDA_delay_equation} gives the pole equation $\det[s\mathbf I+i\Omega\mathbf I+\widetilde{\boldsymbol G}^{\mathrm{inf,LDA}}(s)]=0$ for $\widetilde{\boldsymbol c}_{\mathrm{LDA}}(s)$. By setting $s=-iE$ and using the LDA BIC conditions in Eqs.~\eqref{eq:SM_inf_LDA_BIC_matrix_darkness} and \eqref{eq:SM_inf_LDA_BIC_energy_condition}, the pole equation becomes $\boldsymbol Y^{\mathrm{inf,LDA}}(\bar E_j^{\mathrm{bic}})\bar{\boldsymbol\alpha}_j=\bar E_j^{\mathrm{bic}}\bar{\boldsymbol\alpha}_j$, which is Eq.~\eqref{eq:SM_inf_LDA_energy_equation} evaluated at $E=\bar E_j^{\mathrm{bic}}$.

\subsection{LDA bound-state conditions}
As in the semi-infinite waveguide, any bound state under the LDA is a BIC. Combining Eqs.~\eqref{eq:SM_inf_LDA_spectral_density} and \eqref{eq:SM_inf_LDA_energy_equation} in the infinite waveguide, we obtain the conditions for an LDA BIC with energy $\bar E_j^{\mathrm{bic}}$
\begin{align}
\boldsymbol J^{\mathrm{inf,LDA}}(\bar E_j^{\mathrm{bic}})\bar{\boldsymbol\alpha}_j&=\mathbf0,
\label{eq:SM_inf_LDA_BIC_matrix_darkness}\\
[\bar E_j^{\mathrm{bic}}\mathbf I-\boldsymbol Y^{\mathrm{inf,LDA}}(\bar E_j^{\mathrm{bic}})]\bar{\boldsymbol\alpha}_j&=\mathbf0.
\label{eq:SM_inf_LDA_BIC_energy_condition}
\end{align}
The first condition has the same form as Eq.~\eqref{eq:SM_inf_BIC_darkness}, with $k_j$ and $\alpha_{j\,n}$ replaced by $\bar k_j$ and $\bar\alpha_{j\,n}$. Similar to the semi-infinite waveguide, the exact and LDA BIC conditions are identical for rBICs but generally differ for oBICs.

\subsection{LDA long-time dynamics}
Only the LDA BIC contributions remain in the long-time limit
\begin{equation}
\boldsymbol c_{\mathrm{LDA}}(\infty)=\sum_j\bar{\boldsymbol R}_j^{\mathrm{inf,LDA}}\boldsymbol c(0)e^{-i\bar E_j^{\mathrm{bic}}t},
\label{eq:SM_inf_LDA_long_time}
\end{equation}
where $\bar{\boldsymbol R}_j^{\mathrm{inf,LDA}}$ is the residue matrix at $s=-i\bar E_j^{\mathrm{bic}}$ and has the same form as Eq.~\eqref{eq:SM_collective_LDA_residue} with $\boldsymbol\Delta_{\mathrm{LDA}}(E)$ replaced by $\boldsymbol\Delta^{\mathrm{inf,LDA}}(E)$.

Since the exact and LDA rBIC conditions coincide, we take $\bar{\boldsymbol\alpha}_j=\boldsymbol\alpha_j$. The exact BIC energy condition in Eq.~\eqref{eq:SM_inf_BIC_energy_condition} then gives $\boldsymbol\Delta^{\mathrm{inf}}(\Omega)\boldsymbol\alpha_j=\mathbf0$. Differentiating the exact and LDA expressions for $\boldsymbol\Delta^{\mathrm{inf}}(E)$ and $\boldsymbol\Delta^{\mathrm{inf,LDA}}(E)$ in Eqs.~\eqref{eq:SM_inf_inband_shift} and \eqref{eq:SM_inf_LDA_shift} with respect to $E$ at $E=\Omega$, respectively, and using $\boldsymbol\Delta^{\mathrm{inf}}(\Omega)\boldsymbol\alpha_j=\mathbf0$, we obtain
\begin{equation}
\boldsymbol\alpha_j^\dagger
[\partial_E\boldsymbol\Delta^{\mathrm{inf}}(E)|_{E=\Omega}]
\boldsymbol\alpha_j
=
\boldsymbol\alpha_j^\dagger
[\partial_E\boldsymbol\Delta^{\mathrm{inf,LDA}}(E)|_{E=\Omega}]
\boldsymbol\alpha_j.
\label{eq:SM_inf_rBIC_derivative_preservation}
\end{equation}
Combining Eq.~\eqref{eq:SM_inf_rBIC_derivative_preservation} with the residue expressions in Eqs.~\eqref{eq:SM_collective_residue} and \eqref{eq:SM_collective_LDA_residue}, we obtain $\boldsymbol R_j^{\mathrm{inf,bic}}=\bar{\boldsymbol R}_j^{\mathrm{inf,LDA}}$. Thus the rBIC contributions to the long-time dynamics are exactly preserved under the LDA in the infinite waveguide case, whereas the oBIC contributions generally differ.

\bibliography{SIGA}